\documentclass[floats,floatfix,showpacs,amssymb,prd,twocolumn,superscriptaddress,nofootinbib,nolongbibliography,reprint,prd,aps]{revtex4-2}
\usepackage{textcomp}
\usepackage{amssymb,amsmath,verbatim,mathtools,needspace,enumitem,etoolbox,graphicx,microtype,afterpage,bigints,gensymb,tabularx,xspace,siunitx}
\usepackage[normalem]{ulem}
\usepackage{aasmacros}
\usepackage{multirow}
\usepackage{graphicx}
\usepackage{}
\usepackage{dcolumn}
\usepackage{bm}
\usepackage{lipsum}
\usepackage[usenames, dvipsnames, table]{xcolor}
\usepackage{booktabs}
\usepackage{physics}
\usepackage{soul}
\usepackage{tikz}
\usetikzlibrary{positioning}
\usepackage{algpseudocode}
\usetikzlibrary{arrows.meta, positioning, shapes.multipart}

\usepackage{subcaption}

\AtBeginDocument{\RenewCommandCopy\qty\SI}
\ExplSyntaxOn
\msg_redirect_name:nnn { siunitx } { physics-pkg } { none }
\ExplSyntaxOff
\DeclareSIUnit\solarmass{\ensuremath{M_{\odot}}}
\DeclareSIUnit\parsec{pc}
\DeclareSIUnit\lightspeed{$c$}
\DeclareSIUnit\year{yr}
\DeclareSIUnit\arcsecond{as}
\DeclareSIUnit\astronomicalunit{AU}
\DeclareSIUnit\clight{\ensuremath c}

\definecolor{linkcolor}{rgb}{0.0,0.3,0.5}
\usepackage[colorlinks = true,
            linkcolor = linkcolor,
            urlcolor  = linkcolor,
            citecolor = linkcolor,
            anchorcolor = linkcolor]{hyperref}
\usepackage[capitalize]{cleveref}
\Crefname{equation}{Eq.}{Eqs.}
\Crefname{figure}{Fig.}{Figs.}
\Crefname{tabular}{Tab.}{Tabs.}
\Crefname{section}{Sec.}{Secs.}
\Crefname{subsection}{Sec.}{Secs.}

\usepackage{xcolor}
\definecolor{TealBlue}{HTML}{0072B2}
\definecolor{VioletPurple}{HTML}{CC79A7}
\definecolor{EmeraldGreen}{HTML}{009E73}
\definecolor{GoldenYellow}{HTML}{F0E442}

\usepackage{array}

\renewcommand{\d}{\text{d}}
\usepackage{dsfont}

\usepackage{orcidlink}
\begin{document}
\sisetup{range-phrase=-, range-units=single}

\title{\texttt{samsara}: A Continuous-Time Markov Chain Monte Carlo Sampler for Trans-Dimensional Bayesian Analysis}

\author{Gabriele~Astorino~\orcidlink{0009-0005-1111-6708}}
\email{gabriele.astorino@phd.unipi.it}
\affiliation{Dipartimento di Fisica “E. Fermi”, Università di Pisa, I-56127 Pisa, Italy}
\author{Lorenzo Valbusa Dall'Armi~\orcidlink{0000-0002-0412-8058}}
\email{lorenzo.valbusadallarmi@df.unipi.it}
\affiliation{Dipartimento di Fisica “E. Fermi”, Università di Pisa, I-56127 Pisa, Italy}
\affiliation{INFN, Sezione di Pisa, I-56127 Pisa, Italy}
\author{Riccardo~Buscicchio~\orcidlink{0000-0002-7387-6754}}
\affiliation{Dipartimento di Fisica “G. Occhialini”, Università degli Studi di Milano-Bicocca, Piazza della Scienza 3, 20126 Milano, Italy}
\affiliation{INFN, Sezione di Milano-Bicocca, Piazza della Scienza 3, 20126 Milano, Italy}
\affiliation{Institute for Gravitational Wave Astronomy \& School of Physics and Astronomy, University of Birmingham, Birmingham, B15 2TT, UK}
\author{Joachim~Pomper~\orcidlink{0009-0005-8867-7837}}
\affiliation{Dipartimento di Fisica “E. Fermi”, Università di Pisa, I-56127 Pisa, Italy}
\affiliation{INFN, Sezione di Pisa, I-56127 Pisa, Italy}
\author{Angelo~Ricciardone~\orcidlink{0000-0002-5688-455X}}
\affiliation{Dipartimento di Fisica “E. Fermi”, Università di Pisa, I-56127 Pisa, Italy}
\affiliation{INFN, Sezione di Pisa, I-56127 Pisa, Italy}
\author{Walter~Del~Pozzo~\orcidlink{0000-0003-3978-2030}}
\affiliation{Dipartimento di Fisica “E. Fermi”, Università di Pisa, I-56127 Pisa, Italy}
\affiliation{INFN, Sezione di Pisa, I-56127 Pisa, Italy}

\date{\today}

\begin{abstract}

Bayesian inference requires determining the posterior distribution, a task that becomes particularly challenging when the dimension of the parameter space is large and unknown. This limitation arises in many physics problems, such as Mixture Models (MM) with an unknown number of components or the inference of overlapping signals in noisy data, as in the Laser Interferometer Space Antenna (LISA) Global Fit problem. Traditional approaches, such as product-space methods or Reversible-Jump Markov Chain Monte Carlo (RJMCMC), often face efficiency and convergence limitations. This paper presents \texttt{samsara}, a Continuous-Time Markov Chain Monte Carlo (CTMCMC) framework that models parameter evolution through Poisson-driven birth, death, and mutation processes. \texttt{samsara} is designed to sample models of unknown dimensionality. By requiring detailed balance through adaptive rate definitions, CTMCMC achieves automatic acceptance of trans-dimensional moves and high sampling efficiency. The code features waiting time weighted estimators, optimized memory storage, and a modular design for easy customization. We validate \texttt{samsara} on three benchmark problems: an analytic trans-dimensional distribution, joint inference of sine waves and Lorentzians in time series, and a Gaussian MM with an unknown number of components. In all cases, the code shows excellent agreement with analytical and Nested Sampling results. All these features push \texttt{samsara} as a powerful alternative to RJMCMC for large- and variable-dimensional Bayesian inference problems.

\end{abstract}

\maketitle

\section{Introduction}

Bayesian inference essentially involves the computation of the posterior probability distribution for the quantities of interest, $X$, given a hypothesis, $H$, a logical context $I$, also known as the background information, and some observed data $D$. The entire procedure is encoded in Bayes' theorem:
\begin{equation}
    p(X|D\,H\,I) = \frac{p(X|H\,I)p(D|X\,H\,I)}{p(D|H\,I)}\,.
    \label{eq:Bayes_main}
\end{equation}
Although, at least formally, every inference problem can be solved using Bayes' theorem~\cite{jaynes2003probability}, in practice, the calculations required to determine the posterior distribution $p(X|D\,H\,I)$ can be overwhelmingly challenging. For this reason, most studies dealing with the calculation of $p(X|D\,H\,I)$ rely on numerical methods that generate samples from the posterior distribution. The most widely used and powerful of these methods are Markov Chain Monte Carlo (MCMC) algorithms, in their many variants. In a nutshell, MCMC algorithms are designed to generate samples from a target distribution by stochastically exploring the parameter space. In its simplest implementation, given an initial point for the chain, an MCMC evolves it through a sequence of transitions governed by a user-defined specified proposal distribution.
The evolution is modeled to resemble a diffusion process within the target space, and, thanks to ergodicity, the probability density in any given location is proportional to the time the chain spends there. The key principles behind MCMC algorithms are rooted in statistical mechanics. For instance, a necessary condition for MCMC algorithms to generate samples from the target distribution is that their transitions be stationary, that is, the forward and backward transition probabilities must balance.  In analogy with thermal equilibrium, detailed balance ensures that the process probability distribution remains stationary.

MCMCs are guaranteed to converge to the target distribution\footnote{In other words, the samples collected by the MCMC follow the target distribution.}, given a sufficient, sometimes infinite, number of steps. The successes of MCMC algorithms are ubiquitous, and many branches of modern science would not be as successful without their invention. Nonetheless, designing an MCMC algorithm is far from trivial, particularly when dealing with parameter spaces of large, and possibly unknown, dimensionality.

In essence every MCMC algorithm draws sets of discrete samples from an underlying continuous stochastic process. 
In doing so, it discretizes the ``inference time axis''.
It is useful to discuss this point further, since it plays an important role in the discussion that follows. A standard Metropolis-Hastings (MH) MCMC, used to draw samples from a target distribution $f(x)$ with proposal distribution $q(x|y)$ can be interpreted as follows: 
\begin{quote}
    Given the current state $x_i \equiv x(t_i)$, the next state is the one generated from the proposal distribution $q(x_{i+1}|x_i)$ with a probability determined by the Metropolis-Hastings acceptance ratio \cite{metropolis1953equation,hastings1970monte}.
\end{quote}
The above is equivalent to defining a time axis and checking for transitions to new states at \emph{fixed} sampling intervals, given the state at an earlier time. 

In a continuous-time (CT) representation, an MC aims at simulating a different process: 

\begin{quote}
    Given the current state $x_i$, and a set of $N$ possible states $\{x_j(t_{i+1})\}_{j=1}^{N}$, the next state is $x_k$ with probability $\frac{N!}{x_1! \, x_2! \, \cdots \, x_N!}
\prod_{j=1}^N p_j^{x_j}$, and $p_j = f(x_j)/\sum_jf(x_j)$. 
\end{quote}

Albeit apparently simple, the above interpretation represents a methodological paradigm shift: we represent the inference process as Poisson process, with rates given by our ability of exploring states, given the states we are currently exploring. In other words, the chain is \emph{always} evolving\footnote{In MCMC operative language, the acceptance probability for ``a'' transition is always equal to 1.} toward a more probable state, we just have to be able to find out which one it is. This is analogous to a path integral formulation of the problem of determining the most probable route to go from point A (the prior) to point B (the posterior) in the space of probabilities.

A CT representation of the exploration of parameter spaces also benefits from the theory of stochastic differential equations. A generic continuous-time stochastic process can be described in terms of i) a deterministic drift term, ii) a stochastic diffusion term, and iii) a discontinuous jump term, see \cite{gardiner2004handbook}. 
Therefore, in the CT framework, one can naturally include both random mutations of the states (the diffusion) and changes in the number of parameters necessary to label that state (the jumps)\footnote{We defer to future developments the inclusion of a deterministic drift based on Hamiltonian dynamics, as in HMC-based algorithms \cite{neal2011mcmc}.}. This makes CT-based inference algorithms very appealing for the exploration of parameter spaces of potentially countably infinite dimension.

This paper presents \texttt{samsara}\footnote{In Buddhism, samsara is often defined as the endless cycle of birth, death, and rebirth.}, a Continuous-Time Markov Chain Monte Carlo (CTMCMC) algorithm designed to explore countably infinite-dimensional parameter spaces. Popular examples are Dirichlet and infinite Mixture Models (MM) \cite{gupta2001history,rasmussen2000infinite}, and models of signal inference in noisy data, where the number of signals is a priori unknown.

A widely used approach for handling such spaces is Reversible Jump MCMC (RJMCMC) \cite{Green1995RJMCMC} --- an incarnation of a large class of methods often referred to as \emph{product space methods} \cite{CarlinChib_product_space} --- in which the chain is allowed not only to explore a given parameter space, but also to jump between spaces of different dimensionality -- with an acceptance prescription as for MH ---, thereby exploring models with variable numbers of parameters. 

Our sampler models parameter space exploration as a birth-death-mutation Poisson process, yielding efficient trans-dimensional exploration, whose rates are set by the requirement of detailed balance, while in-dimension exploration relies on conventional MCMC updates.

We present in detail the algorithm behind \texttt{samsara}, we prove the rate prescriptions, and we develop estimators that weight samples by waiting times for improved accuracy. We present the modular implementation and the memory-efficient storage scheme that scales to large populations. We test the code on three benchmark examples: an analytic mixture, a joint recovery of sine waves and Lorentzians time-series, and a Gaussian Mixture model with unknown components, comparing our results with analytical and Nested Sampling outcomes. We show the capabilities of the code and we perform some comparisons about acceptance and efficiency with product-space and RJMCMC approaches.

The paper is organised as follows: in Section~\ref{sec:ctmcmc} we introduce in detail the general theory of CTMCMC and the theoretical choices behind \texttt{samsara}, whose implementation is discussed in Section~\ref{sec:implementation}. In Section~\ref{tests}, we discuss the \texttt{samsara} solution to a few relevant tests cases in general inference. Finally, we conclude with a discussion and a summary of our results in Section~\ref{sec:discussion}. Detailed calculations are presented in Appendices~\ref{app:Rates derivation from detailed balance}, ~\ref{app:Waiting times in CTMCMC}, ~\ref{app:proposals},~\ref{app:diagnostic}, and ~\ref{app:priors}.

\section{Continuous-Time Markov Chain Monte Carlo}
\label{sec:ctmcmc}
\subsection{Trans-dimensional parameter space}
\label{sec:trans-dimensional parameter space}

The CTMCMC algorithm implemented in \texttt{samsara} is designed to perform sampling of trans-dimensional distributions and Bayesian inference on problems involving an unknown number of sources or components of different classes. Among those, the Laser Interferometer Space Antenna (LISA)~\cite{colpi2024lisadefinitionstudyreport} so-called Global Fit ~\cite{Cornish:2005qw,Vallisneri:2008ye} and infinite MM~\cite{gupta2001history,rasmussen2000infinite}. Our algorithm relies on the theory of birth-death processes \cite{Stephens2000BayesianAO, mohammadi2020continuoustimebirthdeathmcmcbayesian}, originally developed to model the biological world and widely used in socio-economics, thus we opted to borrow from their nomenclature. 

We shall refer to the state of the chain --the state $y$ -- as a \textit{society}. Any society is made of a collection of \textit{populations}\footnote{A society with only one species is formally equivalent to a single population.} of \textit{species}-- the classes of competing families of models --, and any population is a collection of \emph{individuals}, ``atomic'' parameter vectors for a given species. Individuals in a species share the same parameter space. We label a species with $\alpha$, the number of parameters of its individuals as $N_{\rm par,\alpha}$, the number of species in a society with $N_{\rm sp}$, the number of individuals in the population of the species $\alpha$ with $N_{\rm pop,\alpha}$. The parameters of the $i$-th individual of the species $\alpha$ are identified for simplicity as $\theta_{i,\alpha}$.

An individual belonging to species $\alpha$ lives in $\Theta_\alpha \subset \mathbb{R}^{N_{\rm par,\alpha}}$, i.e., $\theta_{i,\alpha}\in \Theta_\alpha$. A population $\alpha$ made of $N_{\rm pop,\alpha}$ individuals lives in the Cartesian product of the parameter spaces of the individual sources,
\begin{equation}
    \Omega_{\alpha,N_{\rm pop,\alpha}} = \left( \prod_{k=1}^{N_{\rm pop,\alpha}}\Theta_\alpha \right) \subset \left(\mathbb{R}^{N_{\rm pop,\alpha}}\right)^{N_{\rm par,\alpha}}\, .
    \label{eq:parameter_space_population}
\end{equation}
A society made of $N_{\rm sp}$ species, each comprising $N_{\rm pop,\alpha}$ individuals lives in 
\begin{equation}
    \Lambda_{\{N_{\rm pop,1},\dots,N_{{\rm pop},N_{\rm sp}}\}} = \prod_{\alpha=1}^{N_{\rm sp}} \Omega_{\alpha,N_{\rm pop,\alpha}} \, . 
    \label{eq:parameter_space_society}
\end{equation}

Nevertheless, being the CT Markov Chain formally represented by a point process, we should distinguish between the parameter space in which the full set of parameters lives, and the state space in which the Markov-Chain is represented as a point process.

Let $E_{\alpha, N_{\rm pop, \alpha}}$ be the state space relative to the population $\Omega_{\alpha, N_{\rm pop, \alpha}}$ composed by $N_{\rm pop, \alpha}$ individuals of species $\alpha$, ignoring the labeling of individuals. The full state space is therefore defined as 
\begin{equation}
    E = \prod_\alpha\Bigg(\bigcup_{N_{\rm pop, \alpha}=0}^\infty E_{\alpha,N_{\rm pop, \alpha}}\Bigg) \equiv \prod_\alpha E_\alpha,
    \label{eq:full_space_nolabel}
\end{equation}
where $E_{\alpha,N_{\rm pop, \alpha}}$ are disjoint and $E_\alpha$ is the ‘‘full'' state space for species $\alpha$, i.e., the restriction of $E$ on the subspace relative to species $\alpha$.

As an example, we consider the case of a society, whose individuals represent plane waves.
Each wave is characterized by its amplitude, frequency, and phase, denoted as $\theta_{i,\rm sine}=\begin{pmatrix} A_i, f_i, \phi_i \end{pmatrix}$. The ``sine'' population is therefore the collection of $N_{\rm pop,sine}$ individuals, $\Omega_{\text{sine}, N_{\rm pop, \rm sine}}=\left\{\theta_{1,\rm sine}, \dots, \theta_{N_{\rm pop, sine}}\right\}$. In the case of a single species, the society is uniquely identified by the ‘‘sine population'', $\Lambda = \left\{\Omega_{\rm sine}\right\}$. 
In this case, in each subspace $E_{\text{sine}, N_{\rm pop,sine}}$, the elements are invariant under relabeling of the individuals. For instance, given two fixed sets of parameters $\theta_1$, $\theta_2$, the points $\{\theta_1,\theta_2\}, \{\theta_2,\theta_1\}$ in the space are equivalent in $E_{\rm sine, N_{\rm pop, sine}}$, while in $\Omega_{{\rm sine}, N_{\rm pop,sine}}$ they are distinct vectors.\footnote{For example, in the case in which $\theta$ can assume only two discrete values $a$ and $b$, $\Omega_{\rm sine,2}=\{(a,a),(a,b),(b,a),(b,b)$\}, while $E_{\rm sine, 2}=\{(a,a), (a,b), (b,b)\}.$}

\begin{figure*}
\centering
\includegraphics[width=1.8\columnwidth]{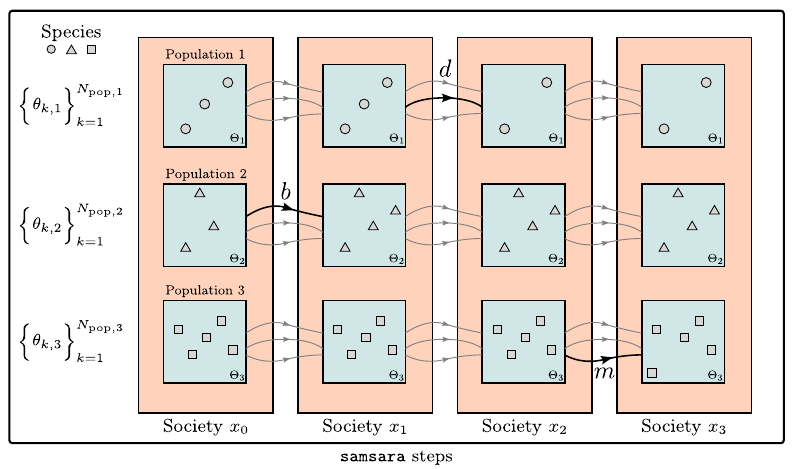}
\caption{
Structure and evolution diagram of \texttt{samsara} trans-dimensional sampling. 
Each step $x_i$ is identified by a society (orange boxes), whose elements are populations (teal boxes) of $N_{\rm pop,\alpha}$ individuals $\theta_{k,\alpha}$, each describing a point in its parameter space $\Theta_\alpha$ of dimension $N_{par,\alpha}$.
Individuals from a population are from the same species $\alpha$. 
For illustrative purposes, we consider the case of three populations from three distinct species. Individuals from each species are denoted with circle, triangle, and square markers. 
At each society evolution step, \texttt{samsara} evaluates all possible birth, death, and mutation rates, following the decision process described in~\Cref{sec:ctmcmc} (grey arrowed lines). 
Then, only one move is chosen and applied to the society, moving onto the next one. We show with a black arrowed line the example of a birth (death, mutation) acting on Population 2 (1, 3), annotated with $b$ ($d$, $m$).    
}
\label{fig:samsara_diagram}
\end{figure*}

\subsection{Birth-death-mutation processes}
\label{sec:moves}

The dynamics of CTMCMC for each species is governed by three processes on $E$: \textit{birth} of a new individual, \textit{death} of an existing individual, and \textit{mutation} of an existing one. In particular, they describe a birth-death process (discontinuous jump term) and a stochastic diffusion process, respectively. We label them with indices $b, d, m$, respectively. For convenience, we define $y_\alpha$ the restriction of the point $y \in E$ to the $E_\alpha$ subspace relative to species $\alpha$, i.e., the point representation of the $\alpha$ population. In what follows, with $y_\alpha \rightarrow y'_\alpha$ we refer to the jump in $E$ given by $y \rightarrow y'=(y\setminus y_\alpha) \cup y'_\alpha$. In full generality, we say that the initial state of the species $\alpha$ contains $N_{\rm pop, \alpha}$ individuals, $y_\alpha = \{\theta_{1,\alpha}, \dots, \theta_{N_{\rm pop, \alpha},\alpha}\}\in E_{\alpha,N_{\rm pop, \alpha}}$. 
\begin{itemize}
    \item \textit{birth}: 
    A birth process for a species $\alpha$ evolves the state by increasing the number of individuals in the population, i.e., it jumps to a final state containing one or more new individuals, $y_\alpha \rightarrow y_\alpha^\prime \in E_{\alpha,N_{\rm pop, \alpha}^\prime}$ with $N_{\rm pop, \alpha}^\prime > N_{\rm pop, \alpha}$.
    In this work, we focus only on the case in which the final state contains the same individuals of the initial one and an additional, $N^\prime_{\rm pop,\alpha}=N_{\rm pop, \alpha} + 1$
    \item \textit{death}: 
    A death process for a species $\alpha$ evolves the state by decreasing the number of individuals in the initial state. The state jumps to one containing one or less individuals, $y_\alpha \rightarrow y_\alpha^\prime \in E_{\alpha,N_{\rm pop, \alpha}^\prime}$ with $N_{\rm pop, \alpha}^\prime < N_{\rm pop, \alpha}$. Again, we focus on the case in which the final state is obtained simply by removing one individual from the initial population,
    \begin{equation}
    \begin{split}
        y'_\alpha & = y_\alpha \setminus \theta_{j, \alpha} \\
        & \equiv \{\theta_{1, \alpha}, \dots, \theta_{j-1, \alpha}, \theta_{j+1, \alpha} \dots, \theta_{N_{\rm pop}, \alpha} \} \, ,
        \label{eq:death_def}
    \end{split}
    \end{equation}
    with $ y'_\alpha \in E_{\alpha, N_{\rm pop,\alpha}-1}$.\\
    
    \item \textit{mutation}: 
    A mutation process for a species $\alpha$ evolves the state by modifying some parameters of some individuals in the population. The final state contains the same individuals, $y_\alpha^\prime \in E_{\alpha,k}$. Hereafter, we will focus only on final states obtained by mutating only a single individual from the initial ones
    \begin{equation}
        y'_\alpha = \{\theta_{1, \alpha}, \dots, \theta_{j-1, \alpha}, \theta_{j, \alpha}', \theta_{j+1, \alpha} \dots, \theta_{N_{\rm pop}, \alpha} \} \, .
        \label{eq:mutation_def}
    \end{equation}
\end{itemize}

Figure \ref{fig:samsara_diagram} shows a schematic representation of the trans-dimensional sampling performed by \texttt{samsara}.

In \texttt{samsara}, for a given species $\alpha$ and process $p$, the algorithm proposes $M_{p,\alpha}$ independent moves to evolve the population. For the death process, we consider $M_{d,\alpha}=N_{\rm pop,\alpha}$ moves, each corresponding to the removal of one of the individuals in the population. For the birth process, multiple individuals could in principle be proposed for the addition to the society.
However, we focus on the case $M_{b,\alpha}=1$. For the mutation process, choosing $M_{m,\alpha}>1$ would correspond to a sampling scheme analogous to the multiple-try proposal~\cite{Liu01032000}. Also in this case, we restrict to $M_{d,\alpha}=1$. We plan to extend to the case $M_{d,\alpha}=1$ in future developments.

In Section~\ref{sec:Dynamics of the Poisson process}, we discuss the procedure to select the species $\alpha$, the process $p$, and the corresponding move $j$ at each step of the MCMC, while a rigorous mathematical description of the moves is provided in Section~\ref{sec:detailed_balance_eq}.

\subsection{Dynamics of the Poisson process}
\label{sec:Dynamics of the Poisson process}

In the CTMCMC framework, birth, death, and mutation are treated at every step as independent Poisson processes, each governed by specific rates, representing the expected number of events occurring per unit time. For the process $p$, the rate associated with the $j$-th of the  $M_{p,\alpha}$ proposed moves for species $\alpha$ is denoted as $R_{j,p,\alpha}$.

For each species, the total rate for a process is given by the sum over the rates of all the proposed moves,
\begin{equation}
    R_{p,\alpha} \equiv \sum_{j=1}^{M_{p,\alpha}} R_{j,p,\alpha}\, . 
    \label{eq:rate_proc_species}
\end{equation}
This rate represents the total number of moves that take place per unit time associated with the process $p$ for each species. Analogously, we define the total rate for a given species as 
\begin{equation}
    R_\alpha = \sum_{p=b,d,m} R_{p,\alpha}\, , 
    \label{eq:rate_species}
\end{equation}
that represents the expected total number of transitions taking place per unit time, evolving the species $\alpha$.

Since births, deaths, and mutations are independent Poisson processes, a society evolves over time as follows.
For each given species, process, and move, we draw the arrival time $t_{j,p,\alpha}$ of the next Poisson count from its exponential distribution with mean $R_{j,p,\alpha}^{-1}$. 
We then evolve the society according to the move with earliest arrival time, i.e. the smallest $t_{j,p,\alpha}$.

Algorithmically, the Poisson-step dynamics implemented in \texttt{samsara} follows a more practical, equivalent route:
\begin{enumerate}
    \item starting from the individual transition rates $R_{j,p,\alpha}$, we evaluate the rates per process of a given species $R_{p,\alpha}$ and the total rates for a species $R_\alpha$ using Eqs.~\eqref{eq:rate_proc_species},~\eqref{eq:rate_species};
    we select the population to evolve by sampling from a categorical distribution with probabilities $p_\alpha = R_\alpha/\sum_{\alpha^\prime} R_{\alpha^\prime}$;
    \item for the species selected, we choose the process by sampling from a categorical distribution with probabilities $p_{p,\alpha} = R_{p,\alpha}/R_\alpha$;
    \item for the species and the process selected, we choose the transition by sampling from a categorical distribution with probabilities $p_{j,p,\alpha} = R_{j,p,\alpha}/R_{p,\alpha}$.
\end{enumerate}
In this workflow, \texttt{samsara} selects the population to evolve according to the total rates associated to each species at the given state.
However, the algorithm also supports a Gibbs-style sampling scheme, as discussed in Section~\ref{sec:detailed_balance_eq}, in which the species of the population to evolve is cycled through deterministically at each MCMC step, while only the process and the specific move are drawn from the rates as described above.

For each state, it is possible to compute an average waiting time, defined as the inverse of the sum of the total rates for the different species, 
\begin{equation}
    \tau = \frac{1}{\sum_\alpha R_\alpha} \, .
    \label{eq:waiting_time}
\end{equation}
The waiting time represents the average lifetime of a state, given the rates $R_{i,p,\alpha}$, i.e., the stability of the state: the larger the waiting time, the larger the probability of such a state. In Section~\ref{sec:Estimators in CTMCMC} we show how the waiting times could be exploited to define more accurate and precise estimators -- such as mean values and covariances -- w.r.t. standard MCMC.

In the next section, we describe how the rates are either computed or fixed, in order to satisfy detailed balance and to guarantee, together with ergodicity, that the trans-dimensional MCMC yields the target distribution as its stationary one.

\subsection{Detailed balance condition}
\label{sec:detailed_balance_eq}

All MCMC algorithms  must satisfy the detailed balance condition and the ergodicity assumption to ensure that the Markov chain admits the target distribution as its stationary one. 

Let $P(\cdot \mid D)$ be\footnote{From now on, we simplify the notation by omitting the hypothesis and the background information in the posterior measure and density.} the target posterior measure with density $p(y|D) = p(y_1 \, \dots \, y_{N_{sp}} | D)$, $y \in E$, and let us define $y_{-\alpha} \equiv y \setminus y_\alpha$. 
Instead of sampling directly from $p(y|D)$, we sample over the species as anticipated in Section \ref{sec:Dynamics of the Poisson process} with two prescriptions. 
The first follows the CTMCMC dynamics, where the population to evolve is selected according to the rates $R_\alpha$ defined in Eq.~\eqref{eq:rate_species}. 
The second approach consists of sampling in turn each species using blocked Gibbs sampling~\cite{geman1984stochastic}, evolving states with a deterministic, predetermined sequence over the species. 
At each step, both approaches require sampling from the conditional distribution $p(y_\alpha | y_{-\alpha} \, D)$\footnote{For simplicity, from now on, we use the compact notation $p(y_\alpha |D) \equiv p(y_\alpha|y_{-\alpha}\, D)$.}, which is, in principle, a trans-dimensional problem in itself. Hence, we sample this conditional distribution using CTMCMC, too. We will see in Section~\ref{algorithm} how they are implemented in practice.

For the first part of this discussion, we focus on birth and death moves, which are each other's inverse process, hence connected through the detailed balance condition. Since the detailed balance for mutations is independent of births and deaths, and it is well known from MCMC theory, we will address it at the end of this section. Assume that a population of species $\alpha$ has been selected for the evolution at a given step of the chain, and denote with $\mu\equiv P(\cdot \mid y_{-\alpha} \, D)$ the target measure with density $p(y_\alpha | D)$, $y_\alpha \in F \subset E_{\alpha,N_{\rm pop,\alpha}}$. 
We also define\footnote{As remarked in Section~\ref{sec:moves}, here we focus exclusively on birth processes that generate only a single new individual. However, in general, $z$ could belong to $E_{\alpha,N_{\rm pop,\alpha}+M}$ with $M$ any positive integer.} $z_\alpha \in G \subset E_{\alpha,N_{\rm pop,\alpha}+1}$ and denote the transition kernels for birth and death by $K_{b,\alpha}
(y_\alpha; G)$,  and  $K_{d,\alpha}
(z_\alpha; F)$ respectively.\footnote{The dimension of the space where they are applied is clear from the point to which they act and the arrival subspace.} More explicitly, they are the probabilities that the state $y_\alpha$ $(z_\alpha)$ transits to a point in the subspace $G$ $(F)$ given the prescription used for the birth (death). 

As in~\Cref{sec:moves}, our birth process involves adding a single individual to the population, proposed from a probability density function $h$ over $\Theta_\alpha$. 
No specific assumptions are imposed on the proposal distribution $h$; it can therefore be chosen freely and may also depend on the current configuration of the individuals in the population.
This is the case for ensemble proposals, where the new point in the MCMC is proposed based in the location of all~\cite{goodman2010ensemble}  -- or a subset~\cite{foremanmackey2013emcee} -- of the others. The birth of an individual $\theta_\alpha$ occurs with rate $R_{b,\alpha}(y, \theta_\alpha)$. The transition kernel from an initial state $y_\alpha$ to the subspace $G$ is the composition of the probability that a birth occurs and the probability that the new point is $\theta_\alpha$, sampled according to $h$, such that the arrival state is in $G$,
\begin{equation}
\begin{split}
    K_{b,\alpha}(y_\alpha; G)\!& =\! P\left(y_\alpha\rightarrow y_\alpha^\prime \in G | \theta_\alpha \right) \, , \\
    y_\alpha^\prime \! &= \! y_\alpha \cup \theta_\alpha \, , \\ 
    \theta_\alpha &\sim h(\cdot \mid y) \, .
\end{split}
\end{equation}

Similarly, our death process consists in the removal of a single individual from the population of species $\alpha$. Each individual in the population, $\theta_{j,\alpha} \in z_\alpha$, dies following a Poisson process, with rate $R_{j,d,\alpha}(z, \theta_{j,\alpha})$, with $j={1, \dots, N_{\rm pop, \alpha}}$. The transition kernel from an initial state $z_\alpha$ to a subspace $F$ is given by the product of the probability of proposing a death move and the probability that the death of an individual yields an arrival state in $F$, summed over all the individuals in the population,
\begin{equation}
    \begin{split}
        K_{d,\alpha}(z_\alpha; F) =& 
        P\left( \bigvee_{\theta_{j, \alpha} \in z_\alpha} z_\alpha \rightarrow z_\alpha \setminus \theta_{j, \alpha}  \in F \right)  \\ \,
        =& \sum_{\theta_{j, \alpha}\in z_\alpha} P(z_\alpha \rightarrow z_\alpha \setminus \theta_{j, \alpha} \in F) \, , 
    \end{split}
\end{equation}
where we used the fact that the union (logical OR) in the first row corresponds to the sum of probabilities, since the events associated with the death of different individuals are mutually exclusive.

The transition kernels reads explicitly
\begin{equation}
    K_{b,\alpha}(y_\alpha; G) = \int_{\Theta_\alpha \, : \,y_\alpha\cup \theta_\alpha\in G} \d \theta_\alpha 
    \frac{R_{b,\alpha}(y, \theta_\alpha)}{R_{b,\alpha}(y)} 
    h(\theta_\alpha|y_\alpha) \, , 
    \label{eq:new_birth_kernel}
\end{equation}
\begin{equation}
    K_{d,\alpha}(z_\alpha; F) = \sum_{\theta_{j, \alpha} \in z_\alpha \, : \, z_\alpha \setminus \theta_{j,\alpha} \in F} \frac{R_{j,d,\alpha}(z, \theta_{j,\alpha})}{R_{d,\alpha}(z)} \, .
    \label{eq:death_kernel}
\end{equation}

According to~\cite{Preston_1975,Stephens2000BayesianAO}, the CT Markov Chain has $p$ as its stationary distribution if the integrated death and birth rates balance
\begin{equation}
    \int_{F}\d \mu(y_\alpha) R_{b,\alpha}(y_\alpha) = \int_{\mathcal{E}_{+1}} \d \mu(z_\alpha) \mathcal{R}_{d,\alpha}(z_\alpha; F) \, ,
    \label{eq:detailed_balance_1}
\end{equation}
\begin{equation}
    \int_{G}\d \mu(z_\alpha) R_{d,\alpha}(z_\alpha) = \int_{\mathcal{E}} \d \mu(y_\alpha) \mathcal{R}_{b,\alpha}(y_\alpha; G) \, .
    \label{eq:detailed_balance_2}
\end{equation}
In Eqs.~\eqref{eq:detailed_balance_1},~\eqref{eq:detailed_balance_2} we have used the compact notation $\mathcal{E}_{+1}$, $\mathcal{E}$ as shorthand for $E_{\alpha, N_{\rm pop,\alpha}+1}, \, E_{\alpha, N_{\rm pop,\alpha}}$, respectively. 
The quantities $\mathcal{R}_{b,\alpha}$, $\mathcal{R}_{d,\alpha}$ represent the transition rate for the birth and death, and are given by
\begin{equation}
    \begin{split}
        \mathcal{R}_{d,\alpha}(z_\alpha; F) \equiv& \,R_{d,\alpha}(z_\alpha)K_{d,\alpha}(z_\alpha; F) \, , \\
        \mathcal{R}_{b,\alpha}(y_\alpha; G) \equiv& \,R_{b,\alpha}(y_\alpha)K_{b,\alpha}(y_\alpha; G) \, .
    \end{split}
\end{equation}

The detailed balance equations establish the relationship between the birth and death rates, the posterior distribution, the birth proposal, and the number of individuals in the population. As a result, we can determine the birth and death rates as a function of the other quantities that appear in the detailed balance conditions, without the need to implement an accept-reject check. Whenever a birth or a death is chosen, given the rates fixed by Eqs~\eqref{eq:detailed_balance_1} and ~\eqref{eq:detailed_balance_2}, the Poisson dynamics, see Section~\ref{sec:Dynamics of the Poisson process}, the proposed point \emph{is always accepted}, so that no calculation is wasted. 

In \texttt{samsara} it is possible to adopt two different prescriptions for the rates calculation. The first, based on~\cite{Stephens2000BayesianAO}, fixes the birth rate $R_{b,\alpha}(y_\alpha)$ and computes the death rates according to 
\begin{equation}
    R_{j, d,\alpha}(y) = \frac{\mathcal{Z}(N_{\rm pop,\alpha}, \theta_{j,\alpha})}{N_{\rm pop, \alpha}}\frac{p(x_\alpha | D)}{p(y_\alpha | D)}R_{b,\alpha}(x_\alpha)h(\theta_{j,\alpha}|x_\alpha) \, , 
    \label{eq:death_rate_fixed_birth}
\end{equation}
where $x_\alpha \equiv y_\alpha \setminus \theta_{j,\alpha}$, and $\mathcal{Z}(N_{\rm pop,\alpha},\theta_{j,\alpha})$ is a factor describing the prior measure. 
The second, proposed in~\cite{mohammadi2020continuoustimebirthdeathmcmcbayesian}, lets both birth and death rates to vary, computing them according to the relations derived from the CTMCMC framework
\begin{equation}
    R_{j, d,\alpha}(y) = \min \left\{1, \frac{\mathcal{Z}(N_{\rm pop,\alpha}, \theta_{j,\alpha})}{N_{\rm pop,\alpha}} \frac{p(x_\alpha | D)}{p(y_\alpha | D)}h(\theta_{j,\alpha}| x_{\alpha}) \right\}\, , 
    \label{eq:death_rate_vary_birth}
\end{equation}
\begin{equation}
    R_{b,\alpha}(y) = \min \left\{1, \frac{N_{\rm pop, \alpha} + 1}{\mathcal{Z}(N_{\rm pop,\alpha}, \theta_{j,\alpha})}\frac{p(z_\alpha | D)}{p(y_\alpha | D)}\frac{1}{h(\theta_{j,\alpha}|y_\alpha)} \right\} \, .
    \label{eq:birth_rate_vary_birth}
\end{equation}
with $z_\alpha \equiv y_\alpha\cup \theta_{j,\alpha}$. We provide mathematical proof for the Eq.~\eqref{eq:death_rate_fixed_birth}, Eq.~\eqref{eq:death_rate_vary_birth} and Eq.~\eqref{eq:birth_rate_vary_birth} and a derivation of the factor $\mathcal{Z}(N_{\rm pop,\alpha}, \theta_{j,\alpha})$ for the test cases discussed in Section~\ref{sec:test_cases} in Appendix~\ref{app:Rates derivation from detailed balance}. 

As anticipated in Section~\ref{sec:moves}, our mutation proposals change only the parameters of a single individual in the society. Mutations are formally defined as transitions with $N_{\rm pop,\alpha}=N^\prime_{\rm pop,\alpha}$. The detailed balance condition for transitions at fixed dimensionality is
\begin{equation}
    p(y_\alpha|D)Q(y_\alpha\rightarrow y_\alpha^\prime) = p(y_\alpha^\prime|D)Q(y_\alpha^\prime\rightarrow y_\alpha)  \, , 
\end{equation}
where the transition probability is given by
\begin{equation}
    Q(y_\alpha\rightarrow y_\alpha^\prime) \equiv \xi(y_\alpha)q(y_\alpha^\prime|y_\alpha) \, ,  
\end{equation}
with $\xi$ the acceptance and $q$ the proposal distribution. For instance, in the MH algorithm, the acceptance is computed using
\begin{equation}
    \xi(y_\alpha) = {\rm min}\left\{1, \frac{p(y_\alpha^\prime|D)}{p(y_\alpha|D)}\frac{q(y_\alpha|y_\alpha^\prime)}{q(y_\alpha^\prime|y_\alpha)}\right\} \, . 
\end{equation}
For fixed-dimensional MCMC samplers, the proposed new point $y_\alpha^\prime$ is accepted with probability $\xi$, while the current state $y_\alpha$ is retained with probability $1-\xi$. Within the Poisson dynamics defined in Section~\ref{sec:Dynamics of the Poisson process}, this corresponds to assigning the following rates for leaving the state $y_\alpha$ or remaining in the current state,
\begin{equation}
    \begin{split}
        R_{m,\alpha}(y_\alpha\rightarrow y_\alpha^\prime) =& R_{m,\alpha}(y_\alpha)\xi(y_\alpha) \, , \\
        R_{m,\alpha}(y_\alpha\rightarrow y_\alpha) =& R_{m,\alpha}(y_\alpha)\left[1-\xi(y_\alpha)\right] \, , 
        \label{eq:mutation_rates}
    \end{split}
\end{equation}
with $R_{m,\alpha}(y_\alpha)$ the total mutation rate per species $\alpha$, Eq.~\eqref{eq:rate_proc_species}. The total mutation rate per species $\alpha$ might depend on the state of the sampler. Hereafter, we fix the rate to a  constant: $R_{m,\alpha}=1$. This allows to design and include mutations based on any MCMC sampling schemes, such as MH variants, ensemble sampling, Hamiltonian Monte Carlo, or any variant of Gibbs sampling.

\subsection{Estimators in CTMCMC}
\label{sec:Estimators in CTMCMC}

An important difference between CTMCMC and standard MCMC is the way they weight states. In CTMCMC each state is associated with a lifetime determined by Poisson dynamics, Section~\ref{sec:Dynamics of the Poisson process}. This section reviews how to compute expectation values in CTMCMC, accounting for unequal and finite lifetimes of the states.

Given a function $f$ of the state $y$, the ergodic theorem ensures that the ensemble average of $f$ can be approximated by averaging $f(y_t)$ along the chain trajectory. In practice, once the chain has reached its stationary distribution, ensemble averages may be replaced by time averages
\begin{equation}
    \left\langle f \right\rangle = \lim_{T\rightarrow \infty}\int_0^T \frac{dt}{T}f\left[y(t)\right] \, . 
\end{equation}
In CTMCMC, the time average is performed using the chain samples according to
\begin{equation}
    \hat{f} = \frac{1}{\sum_{j=1}^{N_{\rm samples}}\Delta t_j}\sum_{i=1}^{N_{\rm samples}}\Delta t_i f\left(y_i\right) \, ,
    \label{eq:estimator_CTMCMC}
\end{equation}
where $\Delta t_i$ is the time that the chain spends in the state $y_i$, which corresponds to the minimum between the $t_{i,p,\alpha}(y_i)$, see Section~\ref{sec:Dynamics of the Poisson process}. Note that, when all $\Delta t_i$ are equal, the estimator defined in Eq.~\eqref{eq:estimator_CTMCMC} is reduces to the standard estimator in equal-time MCMC. In \texttt{samsara}, we do not compute estimates as in Eq.~\eqref{eq:estimator_CTMCMC}, but rather we make use of the Rao-Blackwell theorem~\cite{cf795278-212f-30ee-b4f9-3bf13ee52d9d,Rao1992} and the Rao-Blackwellization scheme described in~\cite{8a19a051-2be1-32f8-8b83-075ff1f85473,LehmannCasella1998}, which was applied in the context of CTMCMC in~\cite{mohammadi2020continuoustimebirthdeathmcmcbayesian}. The Rao-Blackwell theorem ensures that the estimator defined as the expectation value over each $\Delta t_j$ in Eq.~\eqref{eq:estimator_CTMCMC} is an unbiased estimator with lower variance than the original one. The Rao-Blackwellized estimator of $\hat{f}$ is therefore
\begin{equation}
    \hat{f}_{\rm RB} = \frac{1}{\sum_{j=1}^{N_{\rm samples}}\tau_j(y_j)}\sum_{i=1}^{N_{\rm samples}}\tau_i(y_i) f(y_i)  \, ,
    \label{eq:estimator_CTMCMC_RB}
\end{equation}
with $\tau_j$ the waiting times associated with the state $y_j$ defined in Eq.~\eqref{eq:waiting_time}. The Rao-Blackwellized estimator allows us to not sample each time from an exponential distribution, and it could be helpful in reducing the error on estimates of quantities from the CTMCMC. However, the above discussion is valid when the waiting times are representative of all possible transitions in the parameter space given the state in which they are computed. When $M_{b,\alpha}$ are not fixed, the waiting times are related to a transition w.r.t. one state, thus they could fluctuate around their true value. More details on Rao-Blackwellization and the role of waiting times are provided in Appendix~\ref{app:Waiting times in CTMCMC}.

\section{Implementation}
\label{sec:implementation}

\tikzset{
  bigbox/.style={rectangle, draw, rounded corners=4pt, align=center, minimum width=4cm, inner sep=6pt},
  smallcircle/.style={rectangle, draw, align=center, font=\scriptsize, minimum size=0.45cm, inner sep=3pt, line width=0.4pt}
}

\usetikzlibrary{positioning, fit}
\begin{figure*}
\begin{center}
\begin{tikzpicture}[node distance=1.3cm]

\node[bigbox, fill=TealBlue!30] (run) {
  \textbf{run.py}\\[4pt]
  \begin{tikzpicture}[node distance=2pt]
    \node[smallcircle] (n1) {main function};
  \end{tikzpicture}
};

\node[bigbox, fill=TealBlue!30, below=of run] (initialize) {
  \textbf{initialize.py}\\[4pt]
  \begin{tikzpicture}[node distance=2pt]
    \node[smallcircle] (n1) {read the\\ ini file};
    \node[smallcircle, right=of n1] (n2) {initialize the \\ main classes};
  \end{tikzpicture}
};

\node[bigbox, fill=TealBlue!30, below=of initialize] (samsara) {
  \textbf{samsara.py}\\[4pt]
  \begin{tikzpicture}[node distance=2pt]
    \node[smallcircle] (n1) {manage the \\ sampler};
    \node[smallcircle, right=of n1] (n2) {run the \\ sampling};
  \end{tikzpicture}
};

\node[bigbox, fill=EmeraldGreen!30, below=of samsara] (Evolver) {
  \textbf{evolver.py}\\[4pt]
  \begin{tikzpicture}[node distance=2pt]
    \node[smallcircle] (n1) {Evolve the society \\ given the rates};
    \node[smallcircle, right=of n1] (n2) {Store the\\ samples};
  \end{tikzpicture}
};

\node[bigbox, fill=EmeraldGreen!30, below=of Evolver] (rates) {
  \textbf{rates.py}\\[4pt]
  \begin{tikzpicture}[node distance=2pt]
    \node[smallcircle] (n1) {propose\\ new societies};
    \node[smallcircle, right=of n1] (n2) {compute \\ posteriors};
    \node[smallcircle, right=of n2] (n3) {compute \\ rates};
  \end{tikzpicture}
};

\node[bigbox, fill=VioletPurple!30, left=of initialize] (population) {
  \textbf{population.py}\\[4pt]
  \begin{tikzpicture}[node distance=2pt]
    \node[smallcircle] (n1) {individual \\ parameters names};
    \node[smallcircle, right=of n1] (n2) {prior bounds};
    \node[smallcircle, below=of n2] (n3) {signal model};
  \end{tikzpicture}
};

\node[bigbox, fill=GoldenYellow!30, right=of Evolver] (proposal) {
  \textbf{proposal.py}\\[4pt]
  \begin{tikzpicture}[node distance=2pt]
    \node[smallcircle] (nprop1) {propose \\ new points};
    \node[smallcircle, right=of nprop1] (nprop2) {proposal \\ densities};
  \end{tikzpicture}
};

\node[bigbox, fill=VioletPurple!30, left=of Evolver] (posterior) 
{
  \textbf{posterior.py}\\[4pt]
  \begin{tikzpicture}[node distance=2pt]
    \node[smallcircle] (npost1)
    {observations};
    \node[smallcircle, right=of npost1] (npost2)
    {prior};
    \node[smallcircle, below =of npost1] (npost3)
    {likelihood};
  \end{tikzpicture}
};

\node[bigbox, fill=GoldenYellow!30, right=of initialize] (sampler) 
{
  \textbf{sampler.py}\\[4pt]
  \begin{tikzpicture}[node distance=2pt]
    \node[smallcircle] (npost1)
    {MH};
    \node[smallcircle, below=of npost1] (npost2)
    {wrapper for samplers};
    \node[smallcircle, right =of npost1] (npost3)
    {Gibbs};
  \end{tikzpicture}
};

\draw[->] (run) -- (initialize);
\draw[->] (sampler) -- (initialize);
\draw[->] (population) -- (initialize);
\draw[->] (posterior) -- (initialize);
\draw[->] (proposal) -- (initialize);
\draw[->] (initialize) -- (samsara);
\draw[->] (samsara) -- (Evolver);
\draw[->, bend left=25] (Evolver) 
    to node[midway, right] {$N_{\rm gen}$ times} (rates);
\draw[->, bend left=25] (rates) to (Evolver);

\node[draw=TealBlue!200, thick, densely dashed, rounded corners=6pt, inner sep=10pt, fit=(run)(initialize)(samsara),
      label={[font=\bfseries, yshift=0.05cm]above:System control layer}] (controlbox) {};
\node[draw=VioletPurple!200, thick, densely dashed, rounded corners=6pt, inner sep=10pt, fit=(population)(posterior),
      label={[font=\bfseries, yshift=0.05cm]above:Problem specifics (user)}] (controlbox) {};
\node[draw=EmeraldGreen!200, thick, densely dashed, rounded corners=6pt, inner sep=10pt, fit=(Evolver)(rates),
      label={[font=\bfseries, yshift=0.05cm]below:Core engine}] (controlbox) {};
\node[draw=GoldenYellow!200, thick, densely dashed, rounded corners=6pt, inner sep=10pt, fit=(sampler)(proposal),
      label={[font=\bfseries, yshift=0.05cm]above:Sampler specifics (user)}] (controlbox) {};

\end{tikzpicture}
\end{center}

\caption{Code architecture diagram of \texttt{samsara}. The teal blue boxes represent the system control layer of the code, which manages internal processes related to input handling and sampler management. The emerald green boxes are the core engine of the code, related to the Poisson dynamics of the CTMCMC. The violet and golden yellow boxes are the input provided by the user related to the inference problem (population and posterior) and sampler (sampler used and proposals).}
\label{fig:structure_code}
\end{figure*}
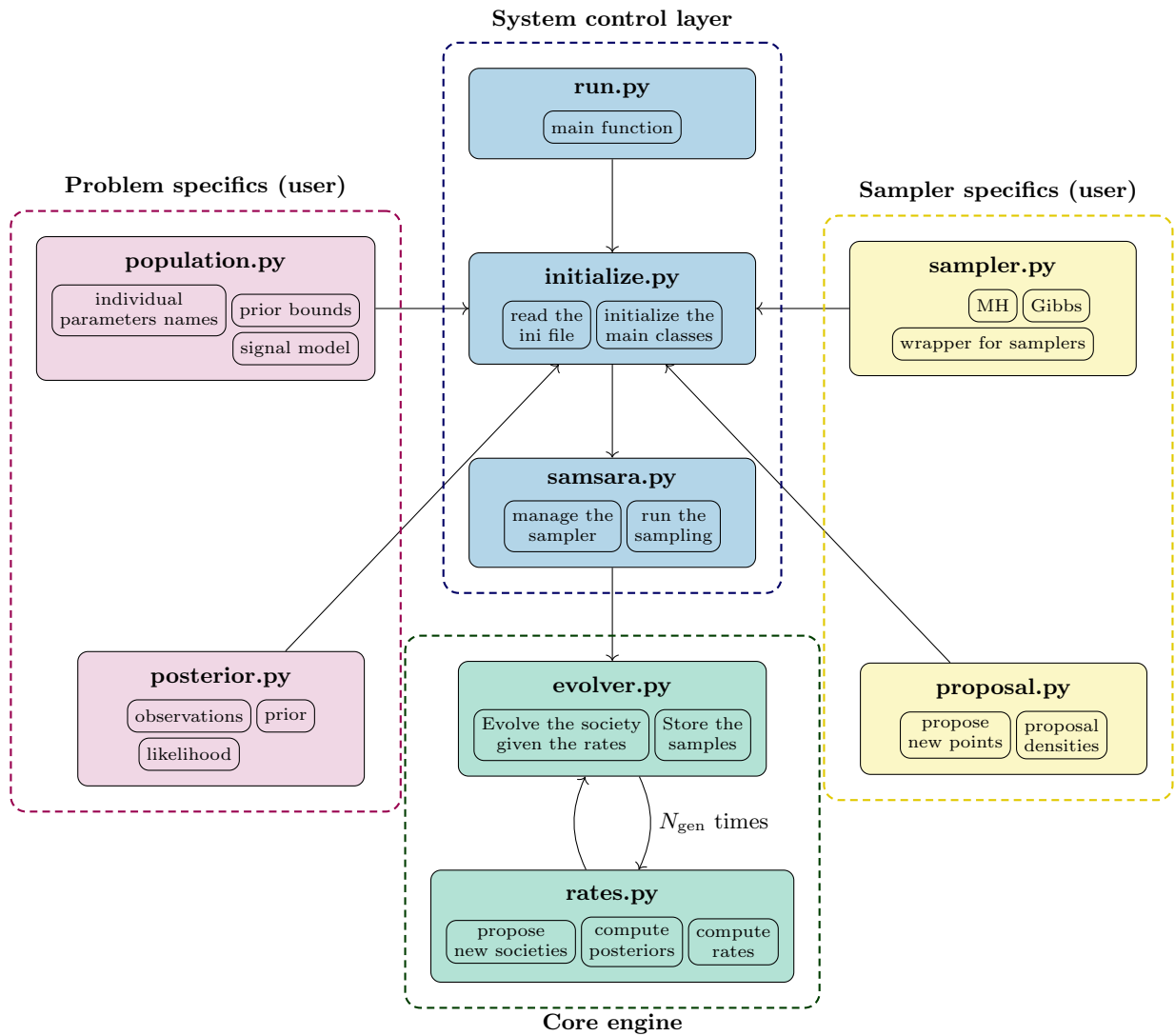

\subsection{Algorithm}\label{algorithm}

\texttt{samsara} samples from trans-dimensional distributions, such as a posterior distribution $p(y|D)$, where the data $D$ have fixed dimensionality, while the parameters $y$ live in the trans-dimensional space described in Section~\ref{eq:parameter_space_society}. For example, $D$ may represent a time or frequency series, and $y$ the set of parameters characterizing the signals present in the data, as in the case of the LISA Global Fit. Alternatively, $D$ could consist of data samples from a certain distribution, and $y$ the parameters of that distribution, as in Gaussian Mixture Models (GMM).

Our algorithm starts by loading the injected data $D$, if the problem requires them, and initializing the initial state of the sampler $y$. This is achieved by defining a society with an initial number of individuals per species, $N_{\rm pop,\alpha}^{\rm ini}$ with parameters $y_\alpha$. The algorithm then proceeds for $N_{\rm gen}$ iterations, repeating the following steps:

\begin{enumerate}
    \item compute the rates $R_{j, p, \alpha}$ for birth and death processes, unless the birth rate is fixed by the user. If the evolution of the population follows the Gibbs-style scheme, the rates are computed only for the population of species $\alpha$ selected to evolve at the current step;
    \label{enum:first_step_algorithm}
    \item determine the species $\alpha^\prime$ of the population to evolve. If the species is selected by the Poisson process, we sample according to $\alpha^\prime \sim \text{Mult}(1|\tau R_1, \, \dots, \, \tau R_{N_{sp}})$, where the total rates per species and the waiting time are defined in Eqs.~\eqref{eq:rate_species},~\eqref{eq:waiting_time}. If the populations evolve via Gibbs-style scheme, $\alpha^\prime$ is chosen deterministically and the code iterates over all the species at each step;
    \item choose the process $p^\prime$ by using the Poisson evolution according to
    \begin{equation}
        p^\prime \sim \text{Mult}\left(\cdot \, \Bigg|1, \, \frac{R_{b,\alpha^\prime}}{R_{\alpha^\prime}}, \, \frac{R_{b,\alpha^\prime}}{R_{\alpha^\prime}}, \, \frac{R_{b,\alpha^\prime}}{R_{\alpha^\prime}}\right) \, ;
    \end{equation}
    \item select the move of a given process according to 
    \begin{equation}
        j^\prime \sim \text{Mult}\left(\cdot \Bigg|1, \, \frac{R_{1,p^\prime,\alpha^\prime}}{R_{p^\prime,\alpha^\prime}}, \, \dots, \, \frac{R_{M_{p^\prime, \alpha^\prime},p^\prime, \alpha^\prime}}{R_{p^\prime,\alpha^\prime}}\right) \, .      
    \end{equation}
    As discussed in Section~\ref{sec:moves}, in the current version of the algorithm, we fix $M_{b,\alpha}=M_{m,\alpha}=1$ and $M_{d,\alpha}=N_{\rm pop,\alpha}$, therefore the selection of the move, i.e., of the individual to remove from the population, is done only for the death;
    \item if $p^\prime$ is a mutation, the new point is accepted according to the prescription described in Section~\ref{sec:detailed_balance_eq}. If $p^\prime=b$ or $p^\prime=d$ the new point is always accepted;
    \item update the society and store it.
\end{enumerate}

\subsection{Overall structure of the code}

$\texttt{samsara}$ is designed as a sampler for a wide range of trans-dimensional problems, such as inference of multiple overlapping signals in noisy data, sampling from trans-dimensional distributions, and mixture models. For this reason, the code is developed with a highly modular architecture, allowing users to implement custom populations, signals models, posteriors, and proposals, without modifying the core of the sampler. 

Figure~\ref{fig:structure_code} illustrates the main modules of the code and briefly outlines their role within the algorithm. In particular, the module \texttt{evolver.py} manages the selection of the species, process, and move used to evolve the society, following the Poisson dynamics described in Section~\ref{sec:Dynamics of the Poisson process}. If the selected process evolving the society is a mutation, the evolution is done by calling the \texttt{run$\_$mcmc} method of the \texttt{sampler} class for the corresponding population.
The module \texttt{rates.py} is used to compute the rates or the acceptance rules derived in Section~\ref{sec:detailed_balance_eq}. 

To implement a new trans-dimensional problem, the user needs to define few functions and classes. For the inference, the user must define a likelihood function, and, optionally, a custom prior, otherwise the code will use a uniform prior for each individual and an unbound improper uniform prior over integer numbers. It is important to stress that as the posterior will always be proper, this presents no difficulties in interpreting the results \cite{bookGelmanBayesianAnalysis3rd}. The prior, along with other attributes, are passed as inputs to the \texttt{posterior} class in the \texttt{posterior.py} module of \texttt{samsara}, while the likelihood must be implemented as a \texttt{posterior}'s method. The user can either use one of the predefined posterior classes provided in the code or define a custom posterior by inheriting from the \texttt{posterior$\_$generator} class in the \texttt{posterior.py} module. The custom likelihood function should take as input a \texttt{society}, defined as a list of \texttt{population} objects defined in the \texttt{population.py} module, one per each species. 

To define a population object, the user must initialize it providing the names of the individual parameters and their corresponding prior bounds. If the individuals contribute to the likelihood through some signals that combine with the data, a \texttt{signal} class must be defined in the \texttt{signals} directory. This class should implement the methods $\texttt{compute\_template}$ and $\texttt{compute\_realization}$, depending on whether the signals are deterministic or stochastic.

The \texttt{data} in the likelihood are passed by the user as an external input and are stored by the algorithm during the run. 

The user can further customize \texttt{samsara} by specifying the proposals for each process and the sampler used for the mutations.  The code currently contains the following proposals: a prior-based proposal and two types of Gaussian proposals—a standard displacements and a mitosis-like proposal\footnote{In biology, mitosis is a process in which a eukaryotic cell divides into two genetically identical daughter cells. The mitosis-like proposal implements this idea by proposing the birth or a mutation of an individual being the copy of one individual from the population, plus a small mutation. See Appendix \ref{app:proposals} for details.}. Alternatively, users can define and implement their own custom proposals. Further details on the available proposals are provided in Appendix \ref{app:proposals}.

In addition, the user can select one of the built-in samplers available in \texttt{samsara} which include the MH or Gibbs and Collapsed Gibbs~\cite{geman1984stochastic} samplers for GMM. Alternatively, the user may implement a wrapper for their own sampler or for any external sampler of choice. \texttt{samsara} already include a wrapper for \texttt{zeus} \cite{karamanis2020ensemble, karamanis2021zeus} and  all built-in sampler, together with the classes required to set up custom wrappers, are contained in \texttt{sampler.py}. 

Finally, the modules \texttt{run.py}, \texttt{initialize.py}, and \texttt{samsara.py} form the internal workflow of the code and are not meant to be accessed or modified by the user. Specifically, \texttt{run.py} runs the CTMCMC algorithm, \texttt{initialize.py} sets up all the quantities specified by the users in the INI files, and \texttt{samsara.py} manages the overall run.

\subsection{Optimized storage of the samples}
\label{sec:StorageSystem}

\begin{table*}[t!]
\renewcommand{\arraystretch}{1.3}
\begin{tabular}{|l|c|c|c|c|c|c|}
\hline
\textbf{Generation} & \textbf{0} & \textbf{1} & \textbf{2} & \textbf{3} & \textbf{4} & \textbf{5} \\
\hline \hline
\textbf{Event} & Empty society & Birth ($\theta_0)$ & Unaccepted drift & Birth ($\theta_1$) & $\theta_0\rightarrow \theta_0^\prime$ & Death ($\theta_1$) \\
\hline 
\textbf{Parameters} 
& $[]$ 
& $[\theta_0]$ 
& $[\theta_0]$ 
& $[\theta_0, \theta_1]$
& $[\theta_0, \theta_1, \theta_0^\prime]$ 
& $[\theta_0, \theta_1, \theta_0^\prime]$ \\ 
\hline
\textbf{Birth Dates} 
& $[]$ 
& $[1]$ 
& $[1]$ 
& $[1, 3]$ 
& $[1, 3, 4]$ 
& $[1, 3, 4]$ \\
\hline
\textbf{Death Dates} 
& $[]$ 
& $[-1]$ 
& $[-1]$ 
& $[-1, -1]$ 
& $[4, -1, -1]$ 
& $[4, 5, -1]$ \\
\hline
\end{tabular}
\caption{Example of the storage indexing system of CTMCMC in the case of one species, starting from an empty society. ‘‘Generation'' represents the step of the chain considered, while ‘‘Event'' the move that takes place at that time.}
\label{tab:IndexingSystem_Example}
\end{table*}

In \texttt{samsara}, we keep track of all individuals in the society in each of the $N_{\rm gen}$ iterations of the sampler. If we assume that each species has an average number of individuals $\bar{N}_\alpha$ during the chain, the memory required to store the full society at each iteration scales as 
\begin{equation}
    \mathcal{M}_{\rm full} \approx N_{\rm gen}\sum_{\alpha=1}^{N_{\rm sp}}\bar{N}_\alpha N_{\rm par}^\alpha \times  8\, \mathrm{B} \, . 
\end{equation}
Such an amount of memory could be very challenging in some cases. For example, in the case of Double White Dwarfs (DWD) systems in LISA~\cite{Cornish:2017vip}, where $\bar{N}_{\rm DWD}\approx 10^3$, $\bar{N}_{\rm par}\approx 10$, and $\bar{N}_{\rm gen}\approx 10^8$, the required memory would be $\mathcal{M}_{\rm full}\approx 10\, \mathrm{TB}$. Furthermore, any kind of post-processing would need to allocate $1-10 \, \mathcal{M}_{\rm full}$ for the time/frequency series of the signals in the Markov Chain.

For this reason, we have optimized the memory used in the algorithm by avoiding redundant copies of the individuals in the society. More specifically, an individual survives in the population either when its proposed drift is not accepted or when other individuals in the society are evolved (through any kind of process). Therefore, the individual's history is fully specified by its birth and death generations. Our algorithm therefore requires just to store the parameters associated with the individual one time and two integers, which refer to the generation of birth, $g_b$, and to the generation of death, $g_d$, of the individual. In this scheme, the memory cost equals the number of unique individuals that appear in the society, multiplied by the memory needed to store their parameters and the two generation indices
\begin{equation}
    \mathcal{M}_{\rm opt} \approx N_{\rm gen}\sum_{\alpha=1}^{N_{\rm sp}}\bar{R}_\alpha^{\rm accept} \,  (N_{\rm par}^\alpha+16) \mathrm{B} \, ,
\end{equation}
with $\bar{R}_\alpha^{\rm accept}$ the average acceptance rate, while the $16\, \mathrm{B}$ are used to store birth and death dates. Since $\bar{R}_\alpha^{\rm accept}\in [0,1]$, $\mathcal{M}_{\rm opt} \leq \mathcal{M}_{\rm full}$ when $\bar{N}_\alpha\geq 3$. Such an indexing storage system is particularly convenient when a large number of individuals in the society are present or when the acceptance rate of the mutation is low.

In \texttt{samsara} all the information is stored in a dictionary $\texttt{saved$\_$samples}$, whose keys correspond to the species in the society. For each species, the value associated with its key is itself a dictionary containing the following entries:
\begin{itemize}
    \item $\texttt{values}$, corresponding to an array of shape $(N_{\rm unique}^\alpha, N_{\rm par}^\alpha)$, with $N_{\rm unique}^\alpha$ the number of unique individuals appeared in the society, i.e., the sum of the number of births for the species $\alpha$ and of the accepted drifts;
    \item $\texttt{lifetime}$, corresponding to an array of shape $(N_{\rm unique}^\alpha, 2)$ where the entry $[i,0]$ and $[i,1]$ are the generations at which the individual with values $i$ entered and left the society, i.e, the birth and death dates.
\end{itemize}
In Table~\ref{tab:IndexingSystem_Example}, we provide an example of how the parameters and the birth and death dates are updated in the algorithm, in the case of one species, starting from an empty population. 

Nevertheless, this specific saving scheme is not always applicable. For instance, in cases where multiple individuals are updated within the same iteration, such as in GMM problems that mutate via Gibbs samplers. For these situations, \texttt{samsara} also provides a standard saving option based on HDF5 (\texttt{.h5}) files to store the samples. The user can freely choose which saving method to employ.

\subsection{Diagnostic}
\label{sec:Diagnostic}

The output of \texttt{samsara} is a chain of $N_{\rm gen}$ samples, in which both the number of individuals per species and the parameter values associated with each individual may vary between samples. The parameters $\theta_{j,\alpha}^{(i)}$ of the $j$-th individual of the $\alpha$-th species of the $i$-th sample form an array of length $N_{\rm par}^\alpha$.

Since the number of individuals per species is not fixed, the correlation length of the chain and other diagnostics cannot be computed as in standard MCMC, because the expectation values of the parameters are ill defined. For this reason, we introduce a scalar function $\rho$ of the society
\begin{equation}
    \rho : \Lambda_{\left\{N_{\rm pop,1}\dots N_{\rm pop,N_{\rm sp}}\right\}} \rightarrow \mathbb{R} \, , 
    \label{eq:scalar_function_society}
\end{equation}
where the parameter space of the society is defined in Eq.~\eqref{eq:parameter_space_society}. The function $\rho$ allows us to extract a scalar quantity from the parameters of the society, bypassing the trans-dimensional nature of the sampling problem. Several choices of $\rho$ are possible, depending on the type of correlation expected in the samples or the diagnostic considered.

For the autocorrelation function we choose $\rho$ to be the log-posterior, as discussed in more detail in Appendix~\ref{app:Correlation_function}, but other choices are possible, such as the number density of the individual species. Similarly, the convergence of the chain is assessed by introducing a set of reference points for each population in the $\Theta_\alpha$ space, and build scalar quantities based on the distance of the population itself from these points~\cite{SissonFan2007, 1992StaSc...7..457G}. Additional details on the distance measures used and the specifics of the convergence tests are given in Appendix~\ref{app:Convergence}.

\subsection{Post-processing}
\label{sec:post-processing}

The post-processing in \texttt{samsara} is done using the thinned samples with correlation length defined in Section~\ref{sec:Diagnostic}. We identify here with $N_{\rm gen}^\prime$ the number of samples after the thinning.

The first quantity we extract is the posterior on the number of individuals per each species in the society, which is obtained by the weighted average of the number of individuals per species, normalized by their mean value,
\begin{equation}
    p\left(N_{\rm pop,\alpha}\right) \equiv \frac{\sum_{i=1}^{N_{\rm gen}^\prime} \tau^{(i)} \boldsymbol{1}_{N_{\rm pop,\alpha}^{(i)}=N_{\rm pop,\alpha}}}{\sum_{i=1}^{N_{\rm gen}} \tau^{(i)}} \, , 
\end{equation}
with $\tau^{(i)}$ the waiting time associated to the $i$-th sample and $\boldsymbol{1}_{\rm condition}$ the indicator function which is one when the condition is true and zero when the condition is false.

The second quantity we extract from the samples is the distribution of the parameters with the population of each species. This quantity could correspond, for instance, to the distribution of angular position in the sky or frequency of DWD. For distribution of the $k$-th parameter of the $\alpha$-th species, we flatten into a single array all the values of the $k$-th parameters from the individuals in the $\alpha$-th population across the $N_{\rm gen}^\prime$ samples. We then plot the normalized histogram of this array, weighting each sample by its corresponding waiting time. We notice that this quantity becomes informative in two regimes. First, when the spacing between source parameters is larger than the uncertainty in the estimates of individual parameters, the resulting distribution reflects the posteriors of individual sources. Second, when the number of sources is large, the distribution reveals the underlying population features, with minimal contamination from the shot noise associated with a single realization.

The post-processing of \texttt{samsara} computes also the median and the confidence intervals of functions of the form
\begin{equation}
    f:\Upsilon \subset \Lambda_{\left\{N_{\rm pop,1}\dots N_{\rm pop,N_{\rm sp}}\right\}} \rightarrow \mathbb{R}^M \, .  
\end{equation}
For example, in the case of GW sources in LISA, \texttt{samsara} can reconstruct the total signal emitted by a given species, such as DWD, in each LISA channel. This is possible because, although the input is trans-dimensional, the output is not - as discussed in Section~\ref{sec:Diagnostic} for the computation of the autocorrelation length.

As already stressed, the output of \texttt{samsara} are $N_{\rm gen}^\prime$ lists of $N_{\rm sp}$ populations, containing $N_{\rm pop,\alpha}$ individuals, whose number can vary between samples. However, one may be interested in getting for each of the $N_{\rm sp}$ species a list of $N_{\rm pop, \alpha}^{\rm eff}$ catalogs of length $N_{{\rm gen}, j,\alpha}$, where each catalog refers to a specific source. Since a catalog should contain at least one sample, the maximum number of catalog one can construct for the species $\alpha$ is $N_{\rm pop,\alpha}^{\rm eff} \leq {\rm max}(N_{\rm pop,\alpha}^{(i)})$. In addition, since some samples could have $N_{\rm pop,\alpha}^{(i)} < N_{\rm pop,\alpha}^{\rm eff}$, the length of the samples for a given catalog is bounded by $N_{{\rm gen},j,\alpha}\leq N_{\rm gen}^\prime$. Schematically, we perform a reordering of the data structure: from a list of samples, each containing a list of populations, each in turn containing a list of individuals, to a list of populations, each containing a list of individuals, each associated with a list of samples. In a coding-style array notation, the unordered and reordered samples can be expressed as
\begin{equation}
    \theta_{j,\alpha}^{(i)} = \begin{cases}\texttt{samples}[i][\alpha][j] \\
    \texttt{catalog}[\alpha][j][i] 
    \end{cases}
\end{equation}
To produce a catalog from the trans-dimensional samples, we use \texttt{PETRA}~\cite{Johnson:2025oyu}, which performs a relabeling of the samples to minimize the statistical divergence between the trans-dimensional samples and a factorized representation of the catalog.

\section{test results}\label{tests}
\label{sec:test_cases}

\subsection{Analytic Test Case}
\label{sec:test_analytic}

\begin{figure}
    \centering
    \includegraphics[width=\columnwidth]{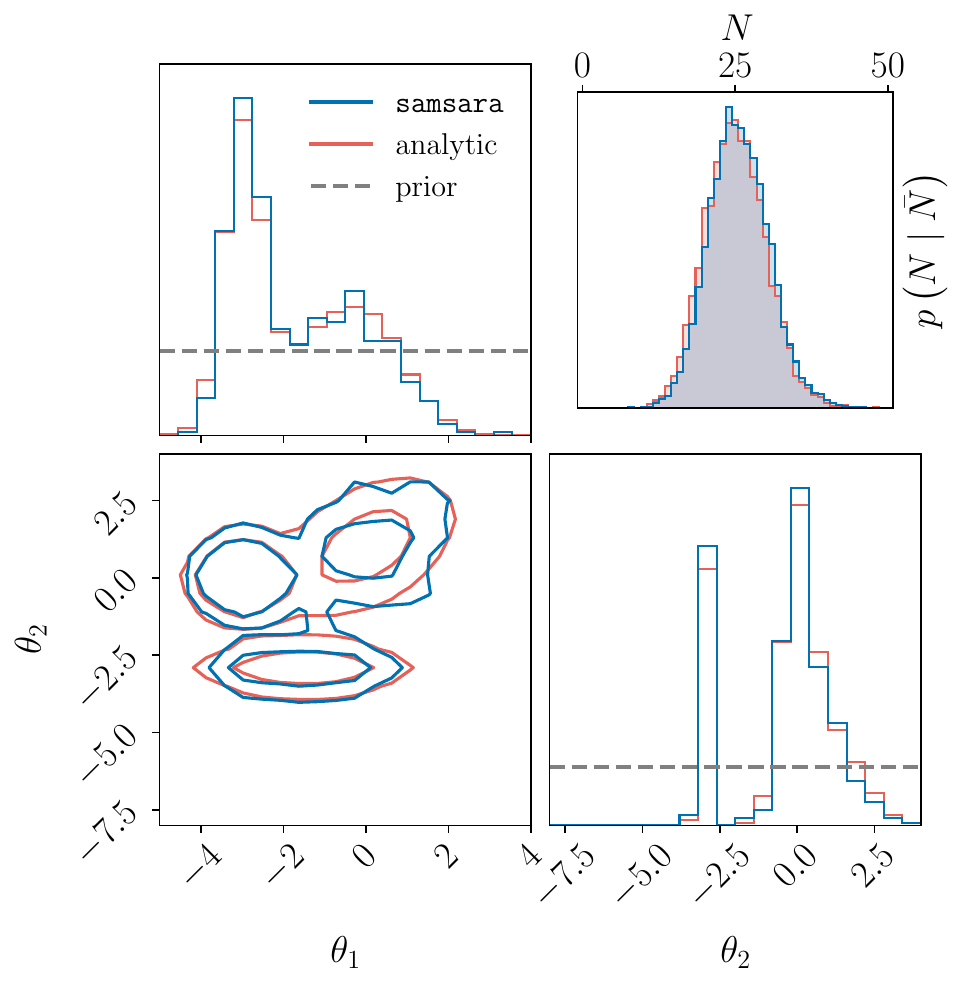}
    \caption{Posterior distribution for the number of sources (upper right panel) and parameter contours (corner plot) for the analytical model described in Eq.~\eqref{eq:analytic_pdf}.
    \texttt{samsara} (analytical) results and priors are shown with deep blue (coral red) and dashed gray lines, respectively. No true values are presented since we are sampling from the trans-dimensional distribution in Eq. \eqref{eq:analytic_pdf} and not performing an inference problem, as discussed in Section~\ref{sec:test_analytic}.}
    \label{fig:analytic_case_posteriors}
\end{figure}

\begin{figure*}[!htb]
    \centering
    \includegraphics[width=1.0\linewidth]{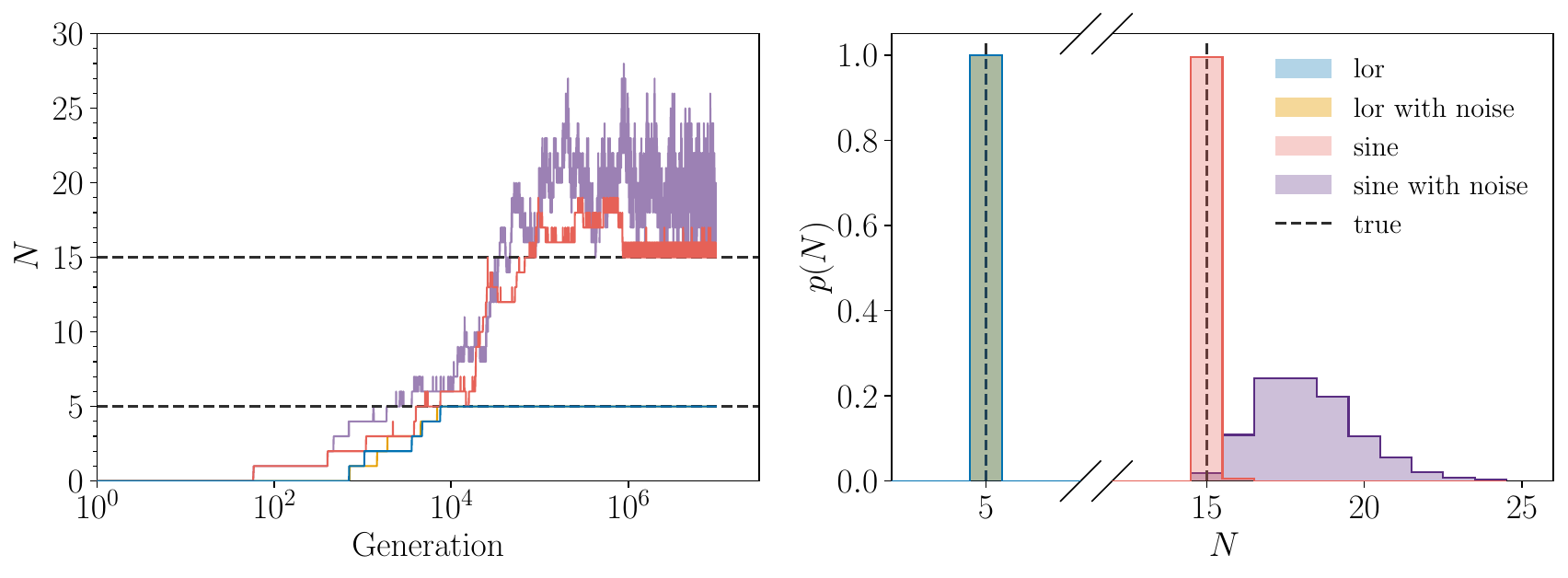}
    \caption{Plot of the trace of the number (left panel) and posterior on the number (right panel). In deep blue (coral red) the Lorentzians (sine waves) for the zero noise case, while in amber (deep purple) the Lorentzians (sine waves) for the noisy case. Truths are in dashed drak gray.}
    \label{fig:sinlines_trace_PofN}
\end{figure*}

As a first check, before addressing more realistic inference problems, we demonstrate that the CTMCMC techniques, used within \texttt{samsara}, faithfully sample trans-dimensional distributions, which in itself is a non-trivial task. This test guarantees that we can use the waiting times $\tau^{(i)}$ to accurately estimate the posterior for the number of signals or the population posterior of the model parameter, as described in section~\ref{sec:post-processing}, for a real inference problem. For this, we replace the likelihood function, dictated by the data, with an analytically understood trans-dimensional distribution 
\begin{equation}
    \begin{split}
    p(N, \theta|\bar{N},\omega) = \mathcal{P}(N|\bar{N})p(\theta|N,\omega) \, , 
    \end{split}
    \label{eq:analytic_pdf}
\end{equation}
and use \texttt{samsara} to sample from it. To mimic a simplified version of the trans-dimensional posterior landscape, as it could be encountered, we choose
\begin{align}\label{eq:theta_giv_N_omega}
        \mathcal{P}(N|\bar{N}) =& \frac{\bar{N}^N}{N!}e^{-\bar{N}} \, , \\
        p(\theta|N, \omega) =& \prod_{i=1}^N p(\theta_i| \omega) \, . \\
        p(\theta_i| \omega) =& \sum_{j=1}^3\frac{w_j}{2\pi\sqrt{{\rm det}\Sigma_j}}e^{-\frac{1}{2}(\theta_i-\mu_j)^T\Sigma_j^{-1}(\theta_i-\mu_j)}
\end{align}
Here, $\mathcal{P}(N|\bar{N})$ represents the posterior of the number of signals via a Poisson distribution, and $p(\theta|N, \omega)$ mimics the population posterior on $\theta$. Therefore, $\lbrace \theta_i\rbrace$ are assumed independently sampled from the population distribution $p(\theta | \omega)$.
Modeling the latter as a Gaussian mixture, we can introduce bi-modalities as well as two probability phases separated by a region of low probability. Such features typically obstruct sampling and allow us to further demonstrate the robustness of the CTMCMC implementation in \texttt{samsara}. Specifically, we choose
\begin{equation}
    \begin{split}
        w =& \left\{
            \frac{8}{18} \, ,  \frac{4}{18} \, ,  \frac{6}{18}
        \right\} \, , \\
        \mu =& \left\{\begin{pmatrix}
        -3 \\ 0
        \end{pmatrix}\, , \begin{pmatrix}
        -1.5 \\ -3
        \end{pmatrix}\, , \begin{pmatrix}
        0 \\ 1
        \end{pmatrix}\right\} \, , \\
        \Sigma = & \left\{\begin{pmatrix}
        0.2 & 0 \\ 0 & 0.2  
        \end{pmatrix} \, , \begin{pmatrix}
        1.3 & 0 \\ 0 & 0.01    
        \end{pmatrix}\, , \begin{pmatrix}
        1 & 0.5 \\ 0.5 & 1    
        \end{pmatrix} \right\} \, .
    \end{split}
\end{equation}
Adopting uniform priors, the replaced likelihood model mimics the full posterior. The priors on $\theta_1$ and $\theta_2$ are listed in Table~\ref{tab:priors}, while for $N$ we adopt the default improper unbound uniform prior over integer numbers. Further, $\mathcal{Z}(N_{\rm pop, \alpha}, \theta_{j,\alpha}) = 1$. The distribution \eqref{eq:analytic_pdf}, while capturing the main features of a trans-dimensional sampling problem, is simple enough to be sampled efficiently by other means and thus allows comparison with the results of our algorithm. 
In Figure~\ref{fig:analytic_case_posteriors}, we show the marginal posteriors for both the number of sources and the population distributions of the model parameters. We compare the post-processed output of \texttt{samsara} with the results of sampling \eqref{eq:analytic_pdf} directly, choosing a number of components and then drawing them independently according to the Gaussian mixture distribution $p(\theta|N, \omega)$ in \eqref{eq:theta_giv_N_omega}. 
For the samples depicted in  Figure~\ref{fig:analytic_case_posteriors} we ran the code for $N_{\rm gen}=10^7$ iterations, with Gaussian proposals for $\theta_1$ and $\theta_2$ with $\sigma_{\rm proposal, \theta_1} = 0.005$, $\sigma_{\rm proposal, \theta_2} = 0.002$. We conclude that the output of \texttt{samsara} is in excellent agreement with the test samples, providing validation for the deployed CTMCMC sampling technique.

Finally, we take advantage of this simple test case to apply the convergence test implemented within \texttt{samsara}. The details are in Appendix \ref{app:Convergence}.

\subsection{Sine waves and Lorentzians}
\label{sec:sinusoids_and_lors}

\begin{figure*}[!htb]
    \centering
    \includegraphics[width=0.48\linewidth]{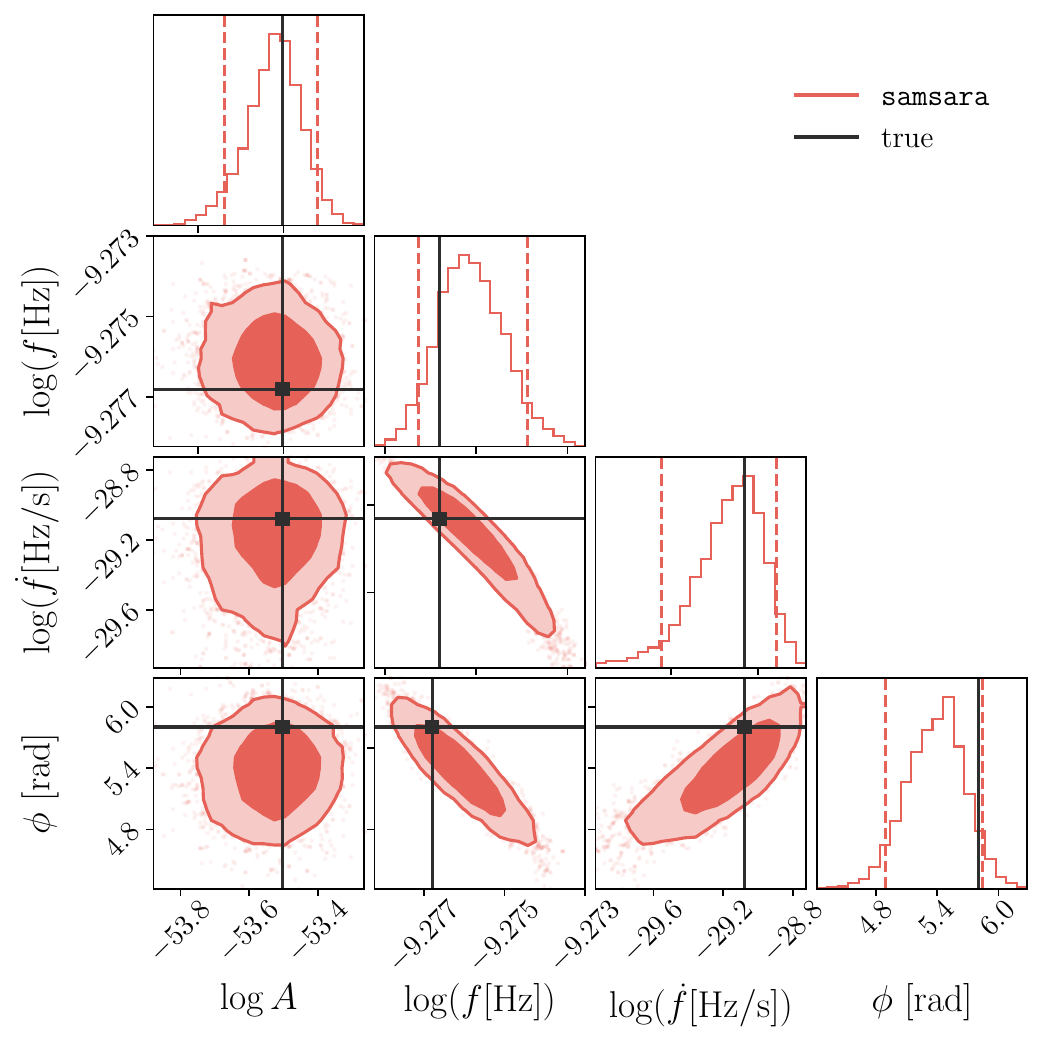}
    \includegraphics[width=0.48\linewidth]{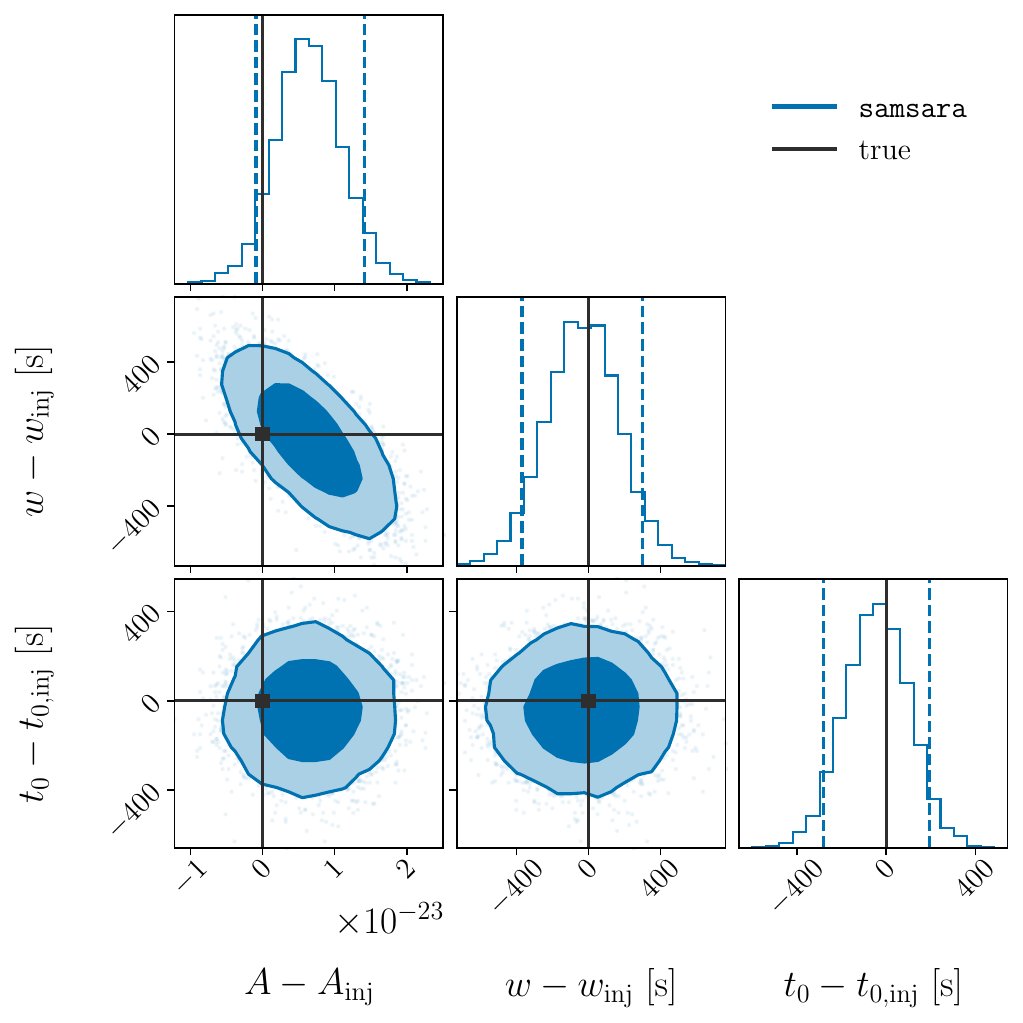}
    \caption{Posterior of one of the injected sine waves (left panel) and one of the injected Lorentzians (right panel). The injected parameters are represented in solid dark gray.}
    \label{fig:sinlines_corners}
\end{figure*}

As a second test, we study the case in which two different kinds of signals may be present in the data. The two species are \textit{sine waves} and \textit{Lorentzians} named ‘‘sine'' and ‘‘lor'', respectively. The sinusoidal signals are described by parameters $\theta_{\rm sine} = \begin{pmatrix} \log A_{\rm sin}, & \log f, & \log \dot{f}, & \phi \end{pmatrix}$, their template is a sine wave with frequency $e^{\log f}$ that evolves in time with constant derivative $e^{\log \dot{f}}$,
\begin{equation}
    s_{\rm sine}(t, \theta_{\rm sine}) = e^{\log A_{\rm sin}}\cos\left(2\pi e^{\log f} t + \pi e^{\log\dot{f}} t^2+\phi \right)\, . 
\end{equation}
The Lorentzians are defined by parameters $\theta_{\rm lor}=\begin{pmatrix} A_{\rm lor}, & w, & t_0\end{pmatrix}$, their template is a Lorentzian centered at time $t_0$ and width $w$,
\begin{equation}
    s_{\rm lor}(t, \theta_{\rm lor}) = A_{\rm lor}\frac{1}{1+\left(\frac{t-t_0}{w}\right)^2}\, .
\end{equation}

As data, we generate a time series with duration $T_{\rm obs}=0.1\, \rm yr$, with cadence for data collection $\Delta t= 500\, s$. In order to resolve the frequency of the simulated signals and observe few modulations of sinusoids, the frequency and its time derivative are constrained by $1/T_{\rm obs} \leq f \leq 1/2\Delta t$, $1/T_{\rm obs}^2 \leq \dot{f} \leq 10/T_{\rm obs}^2$. In order to accurately reconstruct the widths of the Lorentzians, we also impose the condition $w \geq 5 \Delta t$. 

The data are given by
\begin{equation}
    d(t)=\sum_{\alpha=\rm \{sine, lor\}} \sum_{i=1}^{N_{\rm inj,\alpha}} h_{\alpha\,, i}(t)+n(t) \, , 
\end{equation}
where $h_{\alpha\,, i}(t)$ is the signal from the $i-$th injected source of type $\alpha$, with $N_{\rm inj,sine}=15$ and $N_{\rm inj,lor}=5$, and $n(t)$ is the Gaussian noise realization in time domain with covariance $C(t)$. The corresponding log-likelihood of the data is a sum of independent one-dimensional Gaussian terms:
\begin{equation}
    -2\log \mathcal{L} = \sum_t \left\{\frac{\left[d(t)-h(t)\right]^2}{C(t)}+\log\left[2\pi C(t)\right]\right\} \, .
\end{equation}
where $h(t)= \sum_{\alpha,i} h_{\alpha,i}(t)$.
We assume known white noise with constant variance $C(t)=C=10^{-45}$.  

The priors adopted for the parameters of both classes of signals are listed in Table~\ref{tab:priors}. We remark that for both the number of sine waves and Lorentzians we use a uniform prior over all the positive integers. Furthermore, our inference starts from the initial point with $N_{\rm sine}=N_{\rm lor}=0$.

For the mutation, we use adaptive Gaussian proposals, changing the widths accordingly to the signal-to-noise ratio (SNR) of the mutating source. We set $\mathcal{Z}(N_{\rm pop, \alpha}, \theta_{j,\alpha}) = 1$ and $N_{\rm gen}=10^7$, giving a wall time of approximately $15$ hours on a single Intel Xeon Gold core.

\begin{figure*}[!htb]
    \centering
    \includegraphics[width=1.0\linewidth]{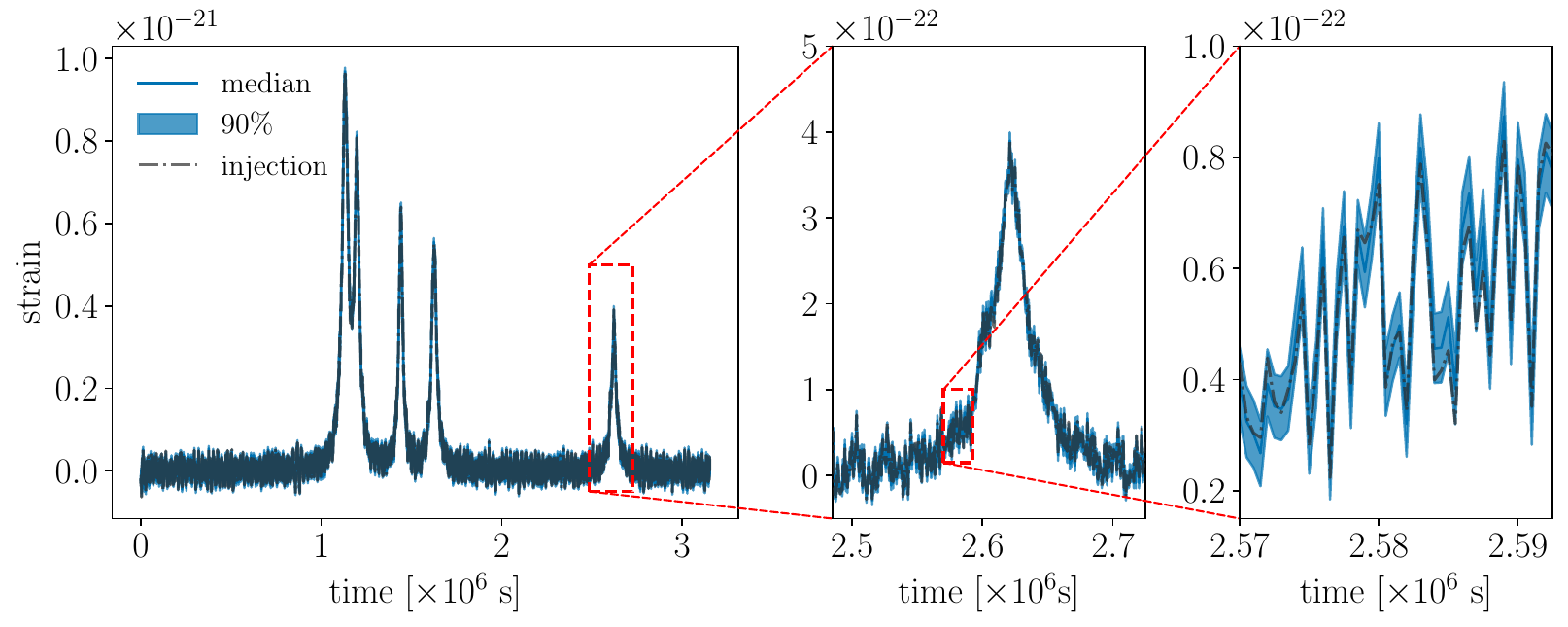}
    \caption{Posterior on the total reconstructed signal (deep blue). The shaded area shows the $90\%$ C.I., while the dash-dot line shows the injected total signal (dark gray).}
    \label{fig:sinlines_signal_posterior}
\end{figure*}

Figure~\ref{fig:sinlines_trace_PofN} shows the so-called trace plots for the number of sources per species as a function of the generations and the posterior on the number performed by \texttt{samsara}. The figure illustrates that the algorithm starts from an initial state with empty society, moves through a low posterior probability region and rapidly converges toward regions of high posterior density, effectively exploring the typical set. In the initial stages of the sampling, \texttt{samsara} typically tends to overfit the total signal by introducing an excess number of sources w.r.t. the true values. These excess sources are progressively removed in the later stages of the evolution. In zero noise, in which we consider the specific realization $n(t)=0$, the numbers of reconstructed sine waves and Lorentzians exactly match the injected values. Conversely, for a generic noise realization, \texttt{samsara} tends to overfit the data with an excess number of sinusoids. This behavior may arise because the sinusoids are a basis, and the effect of the frequency derivative is not strong enough to prevent the algorithm from fitting the spurious contribution due to noise. However, the catalog reconstructed with the \texttt{PETRA} algorithm~\cite{Johnson:2025oyu} described in Section~\ref{sec:post-processing} contains 15 sources (the true number) with a probability of being in the catalog greater than $0.8$ and 10 sources with a probability lower than $0.3$. Therefore, in this case, \texttt{PETRA} is able to distinguish the injected sources from those fitting the noise, which \texttt{samsara} explores through the numerous birth and death moves that make $N_{\rm sine}$ oscillate in Figure~\ref{fig:sinlines_trace_PofN}. In Figure~\ref{fig:sinlines_corners} we show the corner plots of a sine wave (left panel) and a resonance (right panel) reconstructed from the \texttt{samsara} samples using \texttt{PETRA}. All parameters for both signal types have been correctly recovered. In Figure~\ref{fig:sinlines_signal_posterior}, we plot the posterior over the total signal and its 90\% C.I. from the sine wave plus Lorentzians society model. Also in this case, the injected signal is faithfully reconstructed.

\subsection{Gaussian Mixture Model with unknown number of components}
\label{sec:test_GMM}

Finally, we apply \texttt{samsara} to a common problem in inference: clustering of data. Specifically, we focus on the determination of the number of components, and their parameters, in a GMM with an unknown number of components $N$. For the sake of simplicity and ease of comparison, we apply it to the case of uni-variate samples, although \texttt{samsara} can handle arbitrary dimensional cases. We model this problem as a single species society, named \textit{mix}, where an individual is a Gaussian components of the mixture with parameters $\theta_{\rm mix} = \{\pi, \mu, \sigma^2\}$ and density
\begin{equation}
    f_{\rm mix}(x|\theta_{\rm mix}) = \pi\mathcal{N}(x|\mu,\sigma^2) \, .
\end{equation}

We simulate 1000 independent and identically distributed (i.i.d.) samples from the mixture density
\begin{equation}
    p(x|\Vec{\pi}, \Vec{\mu}, \Vec{\sigma}^2) = \sum_{i=1}^3 \pi_i\mathcal{N}(x|\mu_i, \sigma^2_i) \, ,
\end{equation}
with
\begin{equation}
    \begin{split}
        \vec{\pi} =& \left\{\frac{1}{10} \, ,  \frac{6}{10} \, ,  \frac{3}{10} \right\} \, , \\
        \vec{\mu} =& \left\{2, 3, 0\right\} \, , \\
        \vec{\sigma} = & \left\{0.05, 1, 0.4 \right\} \, .
    \end{split}
\end{equation}

We set \texttt{samsara} for GMM problems, in which
\begin{equation}
    \mathcal{Z}(N_{\rm pop,\alpha}, \theta_{j,\alpha}) = \frac{1}{(N_{\rm pop, \alpha} + 1)(1 - \pi_{j,\alpha})^{N_{\rm pop,\alpha}}} \, ,
    \label{eq:factor_rates_GMM}
\end{equation}
and we mutate using the default blocked Gibbs sampler implementation for GMM. In this case, we choose to start with an initial population composed by one individual randomly extracted from the priors. The priors used are discussed again in Appendix \ref{app:priors} and Tabel \ref{tab:priors}.

\begin{figure}[t!]
    \centering
    \includegraphics[width=.9\linewidth]{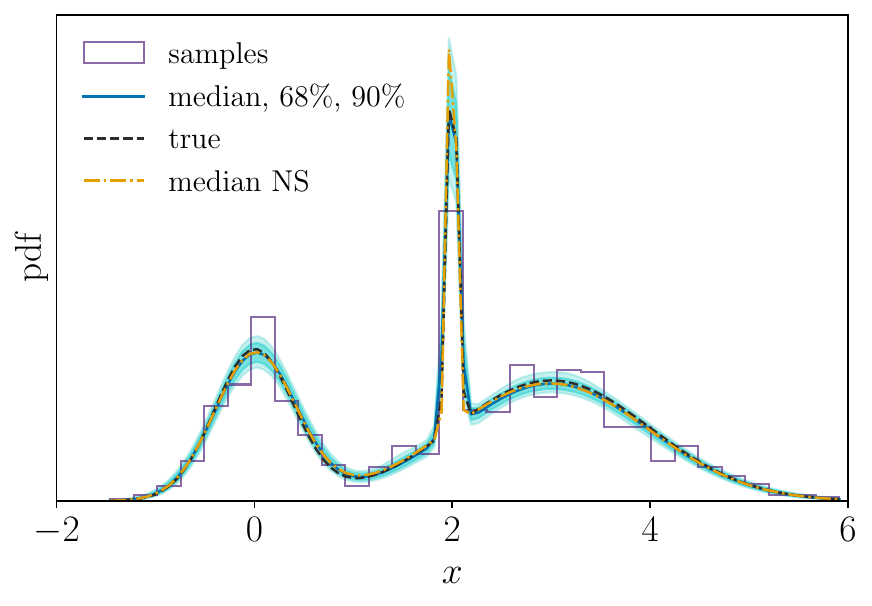}
    \caption{Posterior reconstruction of the GMM both with \texttt{samsara} (median in solid deep blue and $68\%, 90\%$ C.I. in turquoise) and the \texttt{raynest} run corresponding to the maximum of the evidence (median in dashdot amber). Truth mixture model in dashed dark gray. The reconstructed GMMs are essentially indistinguishable.}
    \label{fig:CTMCMC_vs_NS}
\end{figure}

\begin{figure*}[t!]

    \centering
    \includegraphics[width=.9\linewidth]{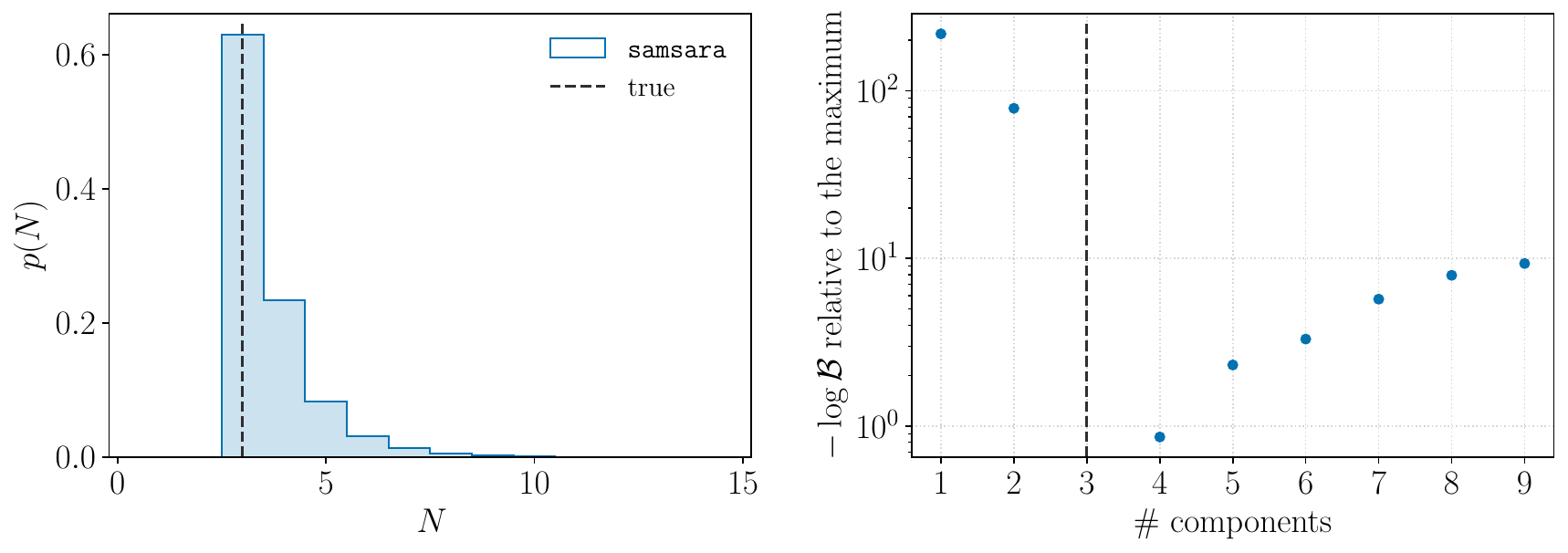}
    \caption{Left panel: posterior distribution on the number of active components from the analysis with \texttt{samsara}. Right panel: negative logarithmic Bayes factor from the nested sampling runs on the product space relative to its maximum value. On both panels, the dashed line indicates the true number of components used to generate the data.}
    \label{fig:GMM_components}
\end{figure*}

We then compare \texttt{samsara} results and performances with the product space prescription using the parallel Python implementation of the nested sampling (NS) \texttt{raynest} \cite{raynest}. The results are summarized in Figures.~\ref{fig:CTMCMC_vs_NS} and \ref{fig:GMM_components}. 

The GMM posterior recovered by \texttt{samsara} accurately describes the samples drawn from the true distribution. Both the true distribution and the median corresponding to the most probable number of components inferred by the NS, lie within the $68\%$ C.R. of the posterior and are fully consistent with the median value. Finally, the posterior on the number of components peaks at three, indicating that this is the most probable number of Gaussian components. Its behavior closely matches that of the Bayes' factor obtained from the product-space procedure.

\section{Discussion}
\label{sec:discussion}

We have presented \texttt{samsara}, a Continuous-Time Markov Chain Monte Carlo sampler for trans-dimensional Bayesian inference. Our framework models the parameter space in terms of a society, which is a collection of populations of different species that are lists of individuals with homogeneous properties, e.g., the number of parameters. The algorithm is based on the Poisson dynamics of birth-death-mutation processes, where species, process, and the transitions used to evolve the society are chosen according to the rates inferred from the proposed new states. A crucial point in our algorithm is that the death and birth rates, unless explicitly fixed by the user, are computed adaptively according to posterior values in the new states. While the mutation rate is fixed, birth and death rates are optimally balanced by the values of the posterior in the new states; the Poisson processes favor a change in the number of individuals in the society only if this provides a substantial gain in the posterior. This feature is particularly important, because, at each step, given the birth-death prescriptions, all the possible transitions are considered and the CTMCMC dynamics moves the chain towards the state of higher probability most of the time.

Our approach presents some advantages w.r.t. RJ-MCMC, where the changes in the number of models used to represent the observations are randomly proposed and accepted with probability given by the acceptance $\xi_{\rm RJ}$, which can be very low. In our framework instead, the exploration of the posterior is self-controlled and optimized at each step by the Poisson process and a birth-death transition is always accepted. The acceptance in CTMCMC $\xi_{\rm CT}$ is typically much larger than $\xi_{\rm RJ}$, because the birth of only a new individual is not affected by the curse of dimensionality.
This means that to reach a state with $N_{\rm ind}$ the number of steps required in CTMCMC are approximately $N_{\rm steps}^{\rm CT} = N_{\rm ind}/ \xi_{\rm CT}$, while for a RJ $N_{\rm steps}^{\rm RJ} = N_{\rm ind}/\xi_{\rm RJ}$ steps. Given that the acceptance of the RJ could be very low, especially for extremely highly-dimensional problems, it is very plausible that $N_{\rm steps}^{\rm RJ} \gg N_{\rm steps}^{\rm CT}$, meaning that we expect our framework to be more efficient in general. This is in stark contrast with the assertions in~\cite{repec:crs:wpaper:2001-29}, where the authors find no efficiency improvements in CTMCMC compared to RJMCMC in very low dimensional problems.

In addition, our algorithm is capable of well exploring the trans-dimensional parameter space not only when starting from a state with a randomly generated society, but even when starting from an empty one. Moreover, it is able to do so with any kind of prior, proper or improper as the default one, being the posterior always proper \cite{bookGelmanBayesianAnalysis3rd}.

\texttt{samsara} is a flexible and highly modular code, designed with a clear separation between the modules responsible for proposing new points, evaluating rates, and evolving the society according to the dynamics of the Poisson process. Moreover, the framework offers users the freedom to fully customize their Bayesian inference setup, by defining their own posterior, specifying the rate prescription, configuring custom proposal mechanisms, or even integrating an external sampler. Alternatively, \texttt{samsara} includes a variety of built-in methods that can be readily used and combined with user-defined components.

We tested the validity of our algorithm with three different illustrative cases that could be relevant both in Bayesian inference, physics, or other research fields. The first case involves sampling from a population whose number of individuals follows a Poisson distribution and whose two-dimensional parameters follow a bi-modal distribution with two probability phases and a sharp peak. The second problem is the joint analysis of sine waves and Lorentzians injected in a noisy time series, where the number and the parameters of both species are unknown. The third one concerns the reconstruction of the probability density function of a Gaussian mixture model with an unknown number of components. In all three cases, \texttt{samsara} successfully samples from the target distribution, accurately recovering both the posterior on the number of components and the posterior on their parameters. These results demonstrate the capability and reliability of \texttt{samsara}, even when compared with Bayes factor statistics obtained from the product-space approach using Nested Sampling.

Nonetheless, several aspects still require improvement and deeper understanding. In particular, we have seen that \texttt{samsara} sometimes can exhibits long autocorrelation lengths, so that we may lose a substantial number of iterations before obtaining statistically independent samples. For instance, this is the case of the test in Section \ref{sec:sinusoids_and_lors}, but it is not for the GMM test of Section \ref{sec:test_GMM} where it is short. We think this arises mostly by the used proposals. Future work will focus on further investigating this issue, and exploring possible improvements which are: investigate the potential advantages or disadvantages of proposing jumps greater than one in the dimensionality of the parameter space, improve \texttt{samsara}'s performances through more sophisticated proposals schemes and more efficient Gibbs and Collapsed Gibbs samplers and study the effect of a deterministic drift for the mutation via HMC. 

\texttt{samsara} is designed to tackle large inference problems in physics and astrophysics. As such, we begun to explore its capabilities to solve the inference over the large number of sources expected in future gravitational waves observatories such as LISA~\cite{colpi2024lisadefinitionstudyreport} and Einstein Telescope~\cite{maggiore2020einstein}. We plan to report our findings in future publications.

\begin{acknowledgments}

The authors are grateful to A. Raimondi for valuable inputs.
L.~V. acknowledges financial support from the project “LISA Global Fit’' funded by the ASI/Università di Trento Grant No. 2024-36-HH.0 - CUP F63C24000390001.
RB acknowledges support from the ICSC National Research Center funded by NextGenerationEU, and the Italian Space Agency grant Phase B2/C activity for LISA mission, Agreement n.2024-NAZ-0102/PER. 

The PhD fellowship of J.P. is funded by the Italian Ministry of University through the project `Nano-Meta-Materials and Devices: New Frontier Concepts for Particle and Radiation Detection' (Grant `Dipartimento di Eccellenza' 2023-2027, CUP I57G22000720004) at the Department of Physics of the University of Pisa.

Computations were performed using University of Birmingham BlueBEAR High Performance Computing facility and CINECA with allocations through INFN, and through EuroHPC Benchmark access call grant EHPC-BEN-2025B08-042.

\noindent\textbf{Data availability statement} The datasets generated during and/or analysed during the current study are available from the corresponding author on reasonable request.\\
\noindent\textbf{Code availability statement} The code/software generated during and/or analysed
during the current study is available from the corresponding author on reasonable request.\\
\noindent\textbf{Open Access} This article is licensed under a Creative Commons Attribution 4.0 International License, which permits use, sharing, adaptation, distribution and reproduction in any medium or format, as long as you give appropriate credit to the original author(s) and the source, provide a link to the Creative Commons licence, and indicate if changes were made. The images or other third party material in this article are included in the article’s Creative Commons licence, unless indicated otherwise in a credit line to the material. If material is not included in the article’s Creative Commons licence and your intended use is not permitted by statutory regulation or exceeds the permitted use, you will need to obtain permission directly from the copyright holder. To view a copy of this licence, visit \href{http://creativecommons.org/licenses/by/4.0/}{http://creativecommons.org/licenses/by/4.0/}.
Funded by $\textrm{SCOAP}^3$.

\end{acknowledgments}

\clearpage

\appendix

\section{Rates derivation from detailed balance}
\label{app:Rates derivation from detailed balance}

In this appendix, we derive Eqs.~\eqref{eq:death_rate_fixed_birth},~\eqref{eq:death_rate_vary_birth},~\eqref{eq:birth_rate_vary_birth} from the detailed balance equations, ensuring together with the ergodicity assumption, that the CT Markov Chain has the target posterior as stationary distribution. We will mainly follow here the computations of~\cite{Stephens2000BayesianAO}.

As shown in Eq.~\eqref{eq:Bayes_main}, Bayes' theorem allows to write the target posterior measure, $P(\cdot \mid D)$, as a function of the likelihood, $P(D|y)$, the evidence, $p(D)$, and the prior measure, $\Pi(\cdot)$,
\begin{equation}
    \d P(y|D) = \frac{p(D|y)}{p(D)}\d \Pi(y) \equiv f(y)\d \Pi(y)\, ,
    \label{eq:bayes_measure_theory}
\end{equation}
where $y$ lives in the space $E$ defined in Section~\ref{sec:trans-dimensional parameter space} and we have defined $f(y)$ as the ratio between the likelihood and evidence.

As discussed in Section~\ref{sec:detailed_balance_eq}, it is possible to sample the posterior from the conditional distribution of a single species $\alpha$. We therefore derive the condition on the rates in this case, fixing the species different from $\alpha$ in the CTMCMC at a given step. For clarity of notation, we use the same conventions of Section~\ref{sec:moves}, defining $y_\alpha\in F\subset E_{\alpha,N_{\rm pop,\alpha}}$ and $y_\alpha^\prime = y_\alpha\cup \theta_\alpha \in G \subset E_{\alpha, N_{\rm pop}+1}$. To highlight the quantities that are conditioned in this computation, we define $y_{-\alpha} \equiv y \setminus y_\alpha$. To simplify the notation we use the compact notation for the restrictions of the posterior and prior measures on $E_{\alpha,N_{\rm pop,\alpha}}$, via
\begin{equation}
    \begin{split}
        \d \mu(y_\alpha) \equiv& \; \d P(y_\alpha|y_{-\alpha} \, D)\, ,  \\
        \d \nu(y_\alpha) \equiv& \; \d \Pi(y_\alpha|y_{-\alpha}) \, .  
    \end{split}
\end{equation}
It is useful to note that Bayes' theorem allows us to write
\begin{equation}
    \int_{\mathcal{E}} \d \mu(y_\alpha) = \int_{\mathcal{E}} \d\nu(y_\alpha) \, f(y|y_{-\alpha})\, ,
    \label{eq:int_post_int_prior}
\end{equation} 
where we have explicitly written that $f$ is conditioned on the value of $y_{-\alpha}$.

For a point process on $E_{\alpha,N_{\rm pop}}$, where the labeling of the individuals is not taken into account, we can write the relation between the prior measure and its density $\pi$ as 
\begin{equation}
    \d \nu(y_\alpha) = \pi(y_\alpha) \d\theta_{1, \alpha} \cdots \d\theta_{N_{\rm pop}, \alpha} \, ,
    \label{eq:prior_no_order}
\end{equation}
where the number dependence is omitted, being clear by $y_\alpha \in E_{\alpha,N_{\rm pop,\alpha}}$.

For what follows, it is useful to note that
\begin{equation}
    \begin{split}
        \pi(y_\alpha)\d \nu(y_\alpha ^\prime) & = \pi(y_\alpha) \pi(y_\alpha^\prime) \prod_{i=1}^{N_{\rm pop,\alpha}}\d\theta_{i, \alpha} \d\theta_{\alpha} \\
        & = \pi(y_\alpha^\prime) \d \nu(y_\alpha)\d\theta_{\alpha} \, . \\
    \end{split}
    \label{eq:priors_relation}
\end{equation}

To recover the rates from the detailed balance equations, we write the left-hand side (LHS) and the right-hand side (RHS) of Eqs.~\eqref{eq:detailed_balance_1},~\eqref{eq:detailed_balance_2} explicitly, substituting the transition kernels of Eqs. \eqref{eq:new_birth_kernel}, \eqref{eq:death_kernel} and integrating on the same integration domain. For simplicity, we focus just on~\eqref{eq:detailed_balance_1}.

The LHS can be written as
\begin{equation}
\begin{split}
    \text{LHS} & = \int_{F}\d \mu(y_\alpha) \, R_{b,\alpha}(y_\alpha)\\ 
    & = \int_{F}\d \nu(y_\alpha) \, f(y_\alpha|y_{-\alpha})R_{b,\alpha}(y_\alpha) \\
    & = \int_{F}\d\nu(y_\alpha) \, f(y_\alpha|y_{-\alpha}) \int_{\Theta_\alpha} \d\theta_{\alpha}^\prime \, R_{b,\alpha}(y, \theta_{\alpha}^\prime) h( \theta_{\alpha}^\prime|y_\alpha)  \\
    & = \int_{F}\int_{\Theta_\alpha} \d\nu(y_\alpha)\d\theta_{\alpha}^\prime \, f(y_\alpha|y_{-\alpha})  \, R_{b,\alpha}(y, \theta_{\alpha}^\prime) h(\theta_{\alpha}^\prime|y_\alpha)\, , 
\end{split}
\end{equation}
The RHS can then be written as
\begin{equation}
\begin{split}
    \text{RHS}& = 
    \int_{\mathcal{E}_{+1}}\d \mu(z_\alpha)\sum_{\theta_{j, \alpha} \in z \ : \ z \ \setminus \ \theta_{j, \alpha} \in F} R_{j,d,\alpha}(z, \theta_{j, \alpha})  \\
    & = \int_{\mathcal{E}_{+1}}\d \mu(z_\alpha)\, \sum_{j=1}^{N_{\rm pop,\alpha}+1} R_{j,d,\alpha}(z, \theta_{j, \alpha})\mathds{1}\{z_\alpha \setminus \theta_{j, \alpha} \in F\} \\
    & = \int_{\mathcal{E}_{+1}}\d\nu(z_\alpha) \, f(z_\alpha|z_{-\alpha}) \\
    & \hspace{6.5em} \sum_{j=1}^{N_{\rm pop,\alpha}+1} R_{j,d,\alpha}(z, \theta_{j, \alpha})\mathds{1}\{z_\alpha \setminus \theta_{j, \alpha} \in F\} \;  \\
    & = \int_{\mathcal{E}_{+1}} \d\nu(z_\alpha) \, (N_{\rm pop, \alpha} + 1) f(z_\alpha|z_{-\alpha}) \\
    &\hspace{7.em} R_{j,d,\alpha}(z, \theta_{j, \alpha})\mathds{1}\{z_\alpha \setminus \theta_{j, \alpha} \in F\} \\
    & = 
    \int_{F} \int_{\Theta_\alpha} \d\nu(y_\alpha) \d \theta_{\alpha}^\prime \, \frac{\pi(y_\alpha^\prime)}{\pi(y_\alpha)} (N_{\rm pop, \alpha} + 1)  \\
    &\hspace{9.em} f(y^\prime|y_{-\alpha}) R_{j,d,\alpha}(y^\prime, \theta_{\alpha}^\prime)\, ,  
    \end{split}
\label{eq:RHS_1_T1}
\end{equation}
with $\mathcal{E}_{+1}$ the compact notation for $E_{\alpha,N_{\rm pop,\alpha}+1}$ introduced in Section~\ref{sec:detailed_balance_eq}, and $y_\alpha^\prime \equiv y_\alpha \cup \theta_\alpha^\prime$. To get the expressions in the second, fourth, and sixth rows, we have used Eq.~\eqref{eq:int_post_int_prior},~\eqref{eq:prior_no_order}, and ~\eqref{eq:priors_relation}. The factor $N_{\rm pop,\alpha}+1$ in the fifth row comes from the symmetry of the prior under the exchange of two individuals. Equating the LHS and RHS, we get 
\begin{equation}
    \begin{split}
        & p(y_\alpha|y_{-\alpha} \, D) R_{b,\alpha}(y, \theta_{\alpha}^\prime) h(y, \theta_{\alpha}^\prime) \\
        &\quad = (N_{\rm pop, \alpha} + 1) p(y_\alpha^\prime | y_{-\alpha} \, D)R_{j,d,\alpha}(y^\prime, \theta_{\alpha}^\prime) \, ,  
    \end{split}
    \label{eq:IL_IR}
\end{equation}
with $R_{b,\alpha}(y, \theta_{\alpha}^\prime), R_{j,d,\alpha}(y^\prime, \theta_{\alpha}^\prime)$ being the unknowns. By repeating the calculation for Eq. \eqref{eq:detailed_balance_2} similarly, we recover again Eq. \eqref{eq:IL_IR}. Say the chain is in state $y$, if we fix $R_{b,\alpha}(y, \theta_{\alpha}^\prime) = R_{b,\alpha}(y_\alpha)$, we have a closed form for $R_{j,d,\alpha}$ and we find Eq. \eqref{eq:death_rate_fixed_birth}. Instead, if both rates are free to vary, we find Eqs.~\eqref{eq:death_rate_vary_birth},~\eqref{eq:birth_rate_vary_birth}.

The factor $\mathcal{Z}(N_{\rm pop,\alpha}, \theta_{j,\alpha})$ introduced in Eq.~\eqref{eq:death_rate_fixed_birth} depends on the relation between the prior measure and the prior density. In this case, where we consider a point process with a prior measure given by Eq.~\eqref{eq:prior_no_order}, we have $\mathcal{Z}(N_{\rm pop,\alpha}, \theta_{j,\alpha})=1$. However, such a factor could change when the label of the individuals has to be taken into account and the parameters in the states are ordered. For example, this applies to GMM, since the weights of the components must live in the simplex. In fact, $N_{\rm pop,\alpha} - 1$ individuals (mixture components) can permute, but the remaining individual is completely determined by the others.

In GMM, we can represent the state as a marked point process~\cite{Stephens2000BayesianAO} on $E$, where the individuals are written as $\theta_\alpha = (w_\alpha, \phi_\alpha)$, with $w_\alpha$ the weight of the component and $\phi_\alpha = \{\mu_\alpha, \Sigma_\alpha\}$ its parameters.

In this case, Eq. \eqref{eq:prior_no_order} becomes
\begin{equation}
\begin{split}
    \d \nu(y_\alpha) & = (N_{\rm pop, \alpha} - 1 )! \, \pi(y_\alpha) \,  \prod_{j=1}^{N_{\rm pop, \alpha}-1} \d w_j \d\phi_{j, \alpha} d\phi_{N_{\rm pop,\alpha}} \, .
\end{split}
\end{equation}

Similarly, we get
\begin{equation}
    \begin{split}
        \d \nu(y_\alpha^\prime) & = \frac{\pi( y_\alpha^\prime)}{\pi(y_\alpha)} N_{\rm pop, \alpha}\d\nu(y_\alpha)\d\phi_\alpha'\d w_\alpha'
    \end{split} \, .
\end{equation}
By repeating the computations done in this appendix keeping into account this extra factor in the prior measure and the fact that the components live in the simplex, we find Eqs.~\eqref{eq:death_rate_fixed_birth},~\eqref{eq:death_rate_vary_birth},~\eqref{eq:birth_rate_vary_birth} with $\mathcal{Z}(N_{\rm pop,\alpha}, \theta_{j,\alpha})$ computed as in Eq. \eqref{eq:factor_rates_GMM}.

\section{Rao-Blackwellization of estimators in CTMCMC}
\label{app:Waiting times in CTMCMC}

\subsection{Rao-Blackwell theorem}

In this appendix, we follow the enunciation of the Rao-Blackwell theorem given in~\cite{LehmannCasella1998}. Let $X$ be a random observable, distributed according to $p_\theta(X=x)$, and be $f$ a function of $\theta$ to estimate. Let then $\delta(X)$ be an estimator for $f$ with finite risk,
\begin{equation}
    R(\theta,\delta) = \mathbb{E}_X\left[L\left(\theta,\delta(X)\right)\right] \, , 
\end{equation}
with $L$ a convex loss function. 

$T$ is said to be a sufficient statistic for $X$ if the conditional distribution of $X$ given $T$ is independent of $\theta$. The Rao-Blackwell theorem states that if $T$ is a sufficient statistic for $p_\theta$, then the Rao-Blackwellized estimator 
\begin{equation}
    \eta(T) \equiv \mathbb{E}_T\left[\delta(X)|T\right]
\end{equation}
satisfies
\begin{equation}
    R(\theta,\eta) \leq R(\theta,\delta) \, , 
\end{equation}
with the inequality saturated if $\delta(X)=\eta(T)$ with probability one. In~\cite{8a19a051-2be1-32f8-8b83-075ff1f85473} the Rao-Blackwellization of estimators has been applied to some sampling schemes, like the Metropolis-Hastings and Accept-Reject algorithm.

\subsection{Rao-Blackwellization with the waiting times}

In CTMCMC, the lifetime of a state is distributed according to an exponential distribution, 
\begin{equation}
    p\left(\Delta t_i|\tau(y_i)\right) = \frac{1}{\tau(y_i)}e^{-\Delta t_i/\tau(y_i)}\, ,
\end{equation}
where we have defined 
\begin{equation}
    \Delta t_i \equiv {\rm min}_{j,p,\alpha} t_{j,p,\alpha} \, , 
\end{equation}
where the $t_{j,p,\alpha}$ are drawn from exponential distributions with average $1/R_{j,p,\alpha}(y_i)$. In this case, $y_i$ is a sufficient statistic for $\Delta t_i$, because the knowledge of $y_i$ completely determines the exponential distribution, therefore the Rao-Blackwellized estimator of Eq.~\eqref{eq:estimator_CTMCMC} is obtained by taking the expectation value of $\Delta t_i$ given $y_i$, i.e., given $\tau(y_i)$, finding Eq.~\eqref{eq:estimator_CTMCMC_RB}.

\section{Proposals}\label{app:proposals}

Currently, \texttt{samsara} includes only a few built-in proposal distributions. Two of them are standard: the prior distribution and a Gaussian displacement. Given the starting individual with parameters $\theta_0$, the prior-based proposal reads
\begin{equation}
    q_{\rm prior}(\theta_t|\theta_0) = \pi(\theta_t) \, ,
    \label{eq:prior_proposal}
\end{equation}
the Gaussian displacement is 
\begin{equation}
    q_{\rm Gaussian}(\theta_t|\theta_0) = \mathcal{N}(\theta_t|\mu=\theta_0,\sigma) \, .
    \label{eq:Gaussian_drift}
\end{equation}

For instance, the default proposal for a birth move is the prior proposal defined in Eq.~\eqref{eq:prior_proposal}. In the GMM case, the default birth proposal is the prior-based following the mutations' Gibbs conjugate priors: Normal-Inverse-Wishart (NIW) for the means and covariances, and Beta distribution for the weight
\begin{equation}
\begin{split}
    h(\mu', \Sigma',\pi'_\alpha) = & \;\text{NIW}(\mu', \Sigma'|\bar{D},k,\text{Cov}(kD),\nu) \times \\
    & \times \text{Beta}(\pi'_\alpha|1,N_{\rm pop, \alpha})\, ,
\end{split}
\end{equation}
where
\begin{equation}
    \begin{split}
        & \text{NIW}(\mu,\Sigma|\mu_0,k,\Lambda,\nu) = \mathcal{N}\left(\mu|\mu_0,\Sigma / k \right)\mathcal{W}^{-1}(\Sigma|\Lambda,\nu)\, , \\
        & 
        \begin{split}
            \mathcal{W}^{-1}(\Sigma|&\Lambda,\nu) = \frac{\det(\Lambda)^{\nu/2}}{2^{\nu M/2}\Gamma_M(\nu/2)} \times \\
            & \times \det(\Sigma)^{-(\nu+M+1)/2}\exp{-\frac{1}{2}\text{tr}\left(\Lambda\Sigma^{-1}\right)}\\
        \end{split} \, , \\
        & \text{Beta}(\pi|\alpha,\beta) = \frac{\Gamma(\alpha+\beta)}{\Gamma(\alpha)\Gamma(\beta)}\pi^{\alpha-1}(1 - \pi)^{\beta-1} \, ,
    \end{split}
\end{equation}
and $\Gamma_M$ the multivariate gamma function.The default choice is $k = 1/5$ and $\nu = M + 2$, with $M$ the dimension of a single data sample in the observations $D$ and $\bar{D}$ is the mean of $D$.

Another possible proposal is more peculiar and genetically inspired. In analogy with biological mitosis, we propose a new individual in the society with parameters $\theta_t$, obtained by applying a multiplicative Gaussian shift to the parameters of another individual $\theta_0$ in the society,
\begin{equation}
    q_{\rm Mitosis}(\theta_t|\theta_0) = \mathcal{N}\left(\theta_t|\theta_0, \xi\theta_0\right) \, . 
\end{equation}
The strength $\xi$ is chosen according to the specific problem considered and may vary across different parameters. In addition, we introduce a \textit{mutation rate} which sets the probability that a component $\theta_t$ remains unaffected by the mutation. In the current version of \texttt{samsara}, the individual $\theta_0$ involved is drawn randomly from the population.

\section{Diagnostic}\label{app:diagnostic}

\subsection{Correlation length}
\label{app:Correlation_function}

According to the discussion in Section~\ref{sec:Diagnostic}, to overcome the issues related to the trans-dimensional nature of the problem, we introduce a scalar function $\rho$ to evaluate the autocorrelation function of the chain. In this work, for simplicity, we consider $\rho$ to be the log-posterior in a given generation, but other choices are possible, such as the number of sources per species. Once the scalar quantity has been chosen, the autocorrelation function is computed according to
\begin{equation}
    {\rm ACF}(\rho,\delta) = \frac{N_{\rm gen}}{N_{\rm gen}-\delta}\frac{\sum_{i=1}^{N_{\rm gen}-\delta}\left(\rho^{(i)}-\bar{\rho}\right)\left(\rho^{(i+\delta)}-\bar{\rho}\right)}{\sum_{i=1}^{N_{\rm gen}}\left(\rho^{(i)}-\bar{\rho}\right)^2} \, , 
    \label{eq:autocorrelation_function}
\end{equation}
where $\bar{\rho}$ is the average value of the scalar function,
\begin{equation}
    \bar{\rho} \equiv \frac{1}{N_{\rm gen}}\sum_{i=1}^{N_{\rm gen}}\rho^{(i)}\, .
\end{equation}
A conservative assumption for the computation of the autocorrelation time is to choose $\delta_{\rm corr}$ as the location of the fifth zero of the autocorrelation function. In Eq.~\eqref{eq:autocorrelation_function} we have neglected the impact of the waiting times introduced in Section~\ref{sec:Estimators in CTMCMC}. Although waiting times are properly taken into account in post-processing, their contribution to the correlation length is negligible. Including them in the computation would generate a nontrivial mode coupling in the correlations at different lags $\delta$, resulting in a complicated expression for the correlation time, which would slow down the code considerably.

\subsection{Convergence}
\label{app:Convergence}

In this section, we discuss the convergence tests within \texttt{samsara}, which are based on~\cite{SissonFan2007, 1992StaSc...7..457G}. 

The problem of quantifying the convergence of trans-dimensional MCMC consists of comparing samples from spaces with different dimensions. This is analogous to the problems encountered in defining an auto-correlation length, as discussed in Section~\ref{sec:Diagnostic}. Also in this case, the idea is to construct an \textit{ad hoc} scalar~\cite{SissonFan2007} used to compare multiple chains of \texttt{samsara}. 

We consider $C$ independent chains, obtained for instance by using different sampler seeds or initial conditions.
Within our framework, for a given species $\alpha$, the $m$-th sample of the trans-dimensional sample path related of the chain $c$, is a population with $N_{\rm pop,\alpha}$ individuals\footnote{Ref.~\cite{SissonFan2007} refers to individuals as components.} denoted by $y_{\alpha,c}^{(m)}$. Each individual has parameters $\theta_{j,\alpha,c}^{(m)} \in \Theta_\alpha \subset \mathbb{R}^{n_\alpha}$. 

We then consider a set of reference points $\{v_{\alpha,\ell} : \ell=1,\dots,N_v, \ v_{\alpha,\ell} \in \mathcal{V}_\alpha \subset \Theta_\alpha\}$, and construct scalar quantities to compare different chains. 
The scalar\footnote{With \emph{scalar}, we mean here a quantity independent on the number of society individuals.} we adopt here is the minimum distance of an individual in the society from a chosen reference point
\begin{equation}
    x_{\alpha, c,\ell}^{(m)} = \min_{i=1,\dots,N_{\rm pop,\alpha}} \left\{ ||\theta_{i,\alpha,c}^{(m)} - v_{\alpha, \ell}|| \right\} \, , 
    \label{eq:B1}
\end{equation}
with $||\cdot||$ the Euclidean norm. Starting from $x_{\alpha, c,\ell}^{(m)}$, it is possible to define the \textit{empirical Cumulative Mass Function} (empirical CMF) for a given reference point $v_{\alpha,\ell}$ as
\begin{equation}
    F^c_{\alpha}(x|v_{\alpha,\ell}) = \frac{1}{N_{\rm gen}^\prime}\sum_{m=1}^{N_{\rm gen}^\prime} \mathds{1}\{x_{\alpha, c,\ell}^{(m)} \le x\} \, ,
    \label{eq:B2}
\end{equation}
where $N_{\rm gen}^\prime$ is the number of thinned chain samples.

With these two quantities three convergence indicators were constructed in  \cite{SissonFan2007}: the \textit{pairwise comparison}, the \textit{Monte Carlo test}, and the \textit{potential scale reduction factor (PSRF)}. 
If $||\cdot||_p$ denotes the $L^p$-norm, the pairwise comparison between two chains reads
\begin{equation}
    u^{c_1c_2}_\alpha \approx \frac{1}{N_v} \sum_{\ell=1}^{N_v}|| F^{c_1}_{\alpha}(x|v_{\alpha,\ell}) - F^{c_2}_{\alpha}(x|v_{\alpha,\ell})||_p \, . 
    \label{eq:B3}
\end{equation}
If $C\ge3$, an overall distance measure can be defined by averaging the pairwise comparison between all the couples, 
\begin{equation}
    u_\alpha = \binom{C}{2}^{-1}\sum_{c_1=1}^C\sum_{c_1\not=c_2} u^{c_1c_2}_\alpha\, .
    \label{eq:B4}
\end{equation}
A convergence criterion for the chains is $u_\alpha\rightarrow0$.
    
For the MC test, we define the quantity
\begin{equation}
    w^c_\alpha \approx \frac{1}{N_v} \sum_{\ell=1}^{N_v} || F^c_{\alpha}(x|v_{\alpha,\ell}) - \bar{F}^c_{\alpha}(x|v_{\alpha,\ell})||_p, \, 
    \label{eq:B5}
\end{equation}
where 
\begin{equation}
    \bar{F}^c_{\alpha}(x|v_{\alpha,\ell}) = \frac{1}{N_{\rm gen}^\prime-1}\sum_{k\not=c}F^k_{\alpha}(x|v_{\alpha,\ell}).
    \label{eq:B6}
\end{equation}
In this case, chains have converged if $w^c_\alpha\rightarrow 0$.
    
The third convergence indicator involves the Gelman-Rubin diagnostic \cite{1992StaSc...7..457G}, using $x_{\alpha, c,\ell}^{(m)}$ as the scalars to be monitored for each $v_{\alpha,\ell}$. In this case, we define the PSRF as 
\begin{equation}
    \hat{R}_{v_{\alpha,\ell}} = \left(\frac{\frac{N_{\rm gen}^\prime-1}{N_{\rm gen}^\prime}W_{v_{\alpha,\ell}} + \frac{1}{N_{\rm gen}^\prime}B_{v_{\alpha,\ell}}}{W_{v_{\alpha,\ell}}}\right)^{1/2}
    \label{eq:B7},
\end{equation}
where 
\begin{equation}
    \begin{split}
        B_{v_{\alpha,\ell}} =& \frac{N_{\rm gen}^\prime}{C-1}\sum_{c=1}^C(\bar{x}_{\alpha, c,\ell} - \bar{x}_{\alpha,\ell})^2, \\
        W_ {v_{\alpha,\ell}}=& \frac{1}{C}\sum_{c=1}^Cs_{\alpha, c, \ell}^2
    \end{split}
\end{equation}
with
\begin{equation}
\begin{split}
    \bar{x}_{\alpha, c,\ell} =& \frac{1}{N_{\rm gen}^\prime}\sum_{m=1}^{N_{\rm gen}^\prime}x_{\alpha, c,\ell}^{(m)}, \\
    \bar{x}_{\alpha,\ell} =& \frac{1}{C}\sum_{c=1}^C\bar{x}_{\alpha, c,\ell} \\
    s_{\alpha, c, \ell}^2=&\frac{1}{N_{\rm gen}^\prime-1}\sum_{m=1}^{N_{\rm gen}^\prime}(x_{\alpha, c,\ell}^{(m)} - \bar{x}_{\cdot c,\ell})^2.
\end{split}
\label{eq:B9}
\end{equation}

For $N_{\rm gen}^\prime\rightarrow\infty$, the convergence is reached in each chain if $\hat{R}_{v_{\alpha,\ell}}\rightarrow1$.

There is no prescription on the number of reference points that should be used for the convergence test. In principle, this choice should depend on the problem considered, but in~\cite{SissonFan2007} $N_v\approx 100$ is suggested as a sufficient number of points. 
For the actual values of points chosen, we randomly extract $N_v$ from points of the different chains.

\subsubsection{Application to the analytic test case}

\begin{figure}
    \centering
   \includegraphics[width=1.\columnwidth]{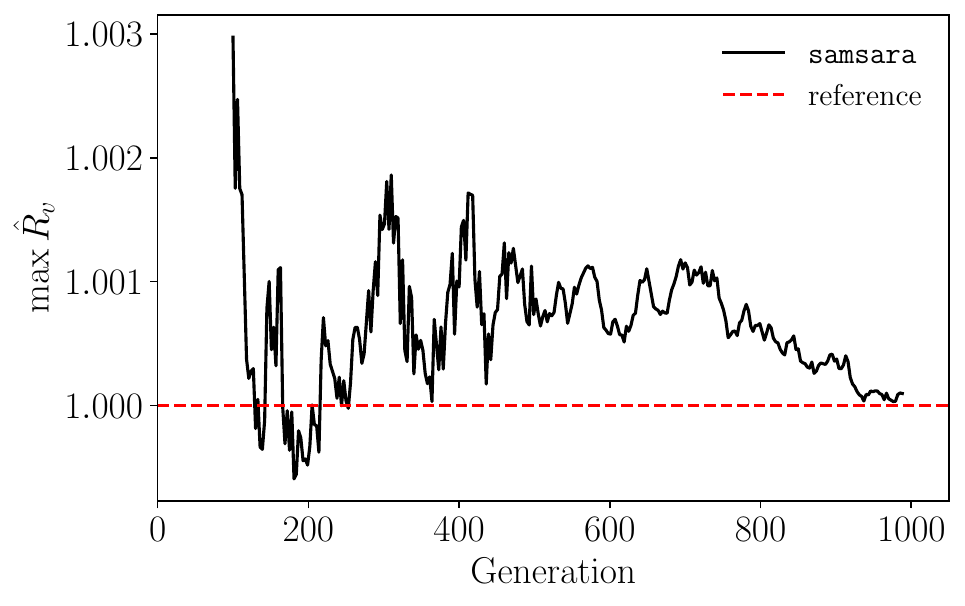}
    \caption{
    Plot of the maximum of the PSRF (solid black line), defined in Eq.~\eqref{eq:B7}, as a function of the generation for the analytic test case discussed in Section~\ref{sec:test_analytic}. The dashed red line shows the reference value of $\hat{R}_v$ used as a criterion for the convergence.}
    \label{fig:convergence_analytic}
\end{figure}

In Figure~\ref{fig:convergence_analytic}, we show the PSRF test applied to the analytic case of Section \ref{sec:test_analytic}. To compute the PSRF defined in Eq.~\eqref{eq:B7}, we collect samples of the distribution~\eqref{eq:analytic_pdf} from 5 independent chains, obtained by initializing \texttt{samsara} with different seeds. After post-processing and thinning the samples as described in Section~\ref{sec:post-processing}, we randomly extract 30 reference points per chain (150 total), computing the PSRF defined in Eq.~\eqref{eq:B7} per each point. To be conservative, we estimate the convergence by looking at the deviation from 1 of the maximum of the PSRF over the ensemble points. Figure~\ref{fig:convergence_analytic} clearly shows that, after the burn-in cut performed, the chains are already consistent with the PSRF test being below 0.3$\%$ deviation from 1.

\begin{table*}[t!]
\centering
\begin{tabular}{l@{\hskip 0.5cm}l@{\hskip 1cm}l}
\toprule
\textbf{Test case} & \textbf{Parameter} & \textbf{Prior} \\
\midrule

\multirow{3}{*}{\textbf{Analytic}} 
  & $N$ & improper unbound uniform on integers \\
  & $\theta_1$ & $\mathcal{U}(-5, 4)$ \\
  & $\theta_2$ & $\mathcal{U}(-8, 4)$ \\
  
  \midrule
  
\multirow{9}{*}{\textbf{Sinusoids and Lorentzians}} 
  & $N_{\rm sine}$ & improper unbound uniform on integers \\
  & $\log A_{\textrm{sine}}$ & $\mathcal{U}(-54.5, -52.5)$ \\
  & $\log (f/\rm Hz)$ & $\mathcal{U}(\log(3\times10^{-5}), \log(10^{-3}))$ \\
  & $\log (\dot{f}/(\rm Hz/s))$ & $\mathcal{U}(\log(10^{-13}), \log(10^{-12}))$ \\
  & $\phi$ & $\mathcal{U}(0, 2\pi)$ \\
  & $N_{\rm lor}$ & $\mathcal{U}(0, \infty)$ \\
  & $A_{\rm lor}$ & $\mathcal{U}(10^{-22}, 10^{-20})$ \\
  & $w \, [s]$ & $\mathcal{U}(10^4, 2\times 10^4)$ \\
  & $t_0\, [s]$ & $\mathcal{U}(0, T_{\rm obs})$ \\[3pt]

  \midrule 

\multirow{3}{*}{\textbf{Gaussian Mixture Model}} 
  & $N$ & improper unbound uniform on integers \\
  & $\pi$ & $\text{Dir}(1/N_{\rm pop,mix},\dots,1/N_{\rm pop,mix})$ \\
  & $\mu$ & $\mathcal{N}(\bar{D},\sigma/k)$ \\
  & $\sigma$ & $\Gamma^{-1}(\text{Cov}(D),\nu)$ \\
  
  \midrule

\bottomrule
\end{tabular}
\caption{Prior range of the parameters for the three different test cases considered in Section~\ref{sec:test_analytic},~\ref{sec:sinusoids_and_lors}, and~\ref{sec:test_GMM}.}
\label{tab:priors}
\end{table*}

\section{Priors}
\label{app:priors}

In all the test cases discussed in Sections~\ref{sec:test_analytic},~\ref{sec:sinusoids_and_lors}, and~\ref{sec:test_GMM}, we adopted an implicit unbounded prior over the number, with each number being equally-likely. Hence, we are fully agnostic on the number of sources, and the algorithm is free to explore any possible dimension of the parameter space. This emphasizes the efficiency of our algorithm in reconstructing the true number of individuals. In Table~\ref{tab:priors}, we list all the priors used in this work. 

For the analytic case, we choose a uniform prior in $\theta_1$, $\theta_2$ large enough to contain all the Gaussian distributions in the mixture.

For the sine waves, in order to resolve the frequency and observe few modulations of sinusoids, the frequency and its time derivative are constrained by $1/T_{\rm obs}\leq f \leq 1/2\Delta t$ and $1/T_{\rm obs}^2 \leq \dot{f} \leq 10/T_{\rm obs}^2$. For the Lorentzians, the prior bounds are chosen such that the peaks lie within the observation window and $w \geq 5 \Delta t$, ensuring that each bump includes a sufficient number of sampling points.

For the GMM, we maintain faith to the blocked Gibbs sampler for the mutation using the Normal-Inverse-Gamma (NIG) for means and covariances, and the Dirichlet distribution (Dir) for the weights, being conjugate priors. Here, the NIG is the one-dimensional equivalent of the Normal-Inverse-Wishart (NIW), 
\begin{equation}
    \begin{split}
        \Gamma^{-1}(\sigma^2|a,b) = \frac{b^a}{\Gamma(a)}\frac{e^{-b/\sigma^2}}{(\sigma^2)^{a+1}}\, ,
    \end{split}
\end{equation}
with $\Gamma$ the gamma function. We set the NIG hyperparameters to be their default initialization of the Gibbs sampler implemented in \texttt{samsara} as already presented in Section \ref{app:proposals}. We set the concentration parameter of the Dirichlet to be $1/N_{\rm pop,mix}$ for each weight, with $N_{\rm pop,mix}$ the number of individuals (components) in the mixture.

\bibliographystyle{apsrev4-2}
\bibliography{tech.bib}

@article{Preston_1975, 
    title={Spatial birth and death processes}, volume={7}, 
    DOI={10.1017/S0001867800040726}, 
    number={3}, 
    journal={Advances in Applied Probability}, author={Preston, Chris}, 
    year={1975}, 
    pages={465–466}
}

@article{Stephens2000BayesianAO,
  title={Bayesian analysis of mixture models with an unknown number of components- an alternative to reversible jump methods},
  author={Matthew Stephens},
  journal={Annals of Statistics},
  year={2000},
  volume={28},
  pages={40-74},
  url={https://api.semanticscholar.org/CorpusID:22613572}
}

@article{Johnson:2025oyu,
    author = "Johnson, Aaron D. and Roulet, Javier and Chatziioannou, Katerina and Vallisneri, Michele and Trejo, Chris G. and Gersbach, Kyle A.",
    title = "{From the LISA global fit to a catalog of Galactic binaries}",
    eprint = "2502.14818",
    archivePrefix = "arXiv",
    primaryClass = "gr-qc",
    doi = "10.1103/95c5-sblc",
    journal = "Phys. Rev. D",
    volume = "112",
    number = "2",
    pages = "024045",
    year = "2025"
}

@article{Liu01032000,
author = {Jun S. Liu and Faming Liang and Wing Hung Wong},
title = {The Multiple-Try Method and Local Optimization in Metropolis Sampling},
journal = {Journal of the American Statistical Association},
volume = {95},
number = {449},
pages = {121--134},
year = {2000},
publisher = {ASA Website},
doi = {10.1080/01621459.2000.10473908},
URL = {https://www.tandfonline.com/doi/abs/10.1080/01621459.2000.10473908},
}

@misc{mohammadi2020continuoustimebirthdeathmcmcbayesian,
      title={Continuous-Time Birth-Death MCMC for Bayesian Regression Tree Models}, 
      author={Reza Mohammadi and Matthew Pratola and Maurits Kaptein},
      year={2020},
      eprint={1904.09339},
      archivePrefix={arXiv},
      primaryClass={stat.ML},
      url={https://arxiv.org/abs/1904.09339}, 
}

@article{cf795278-212f-30ee-b4f9-3bf13ee52d9d,
 ISSN = {00034851},
 URL = {http://www.jstor.org/stable/2236107},
 author = {David Blackwell},
 journal = {The Annals of Mathematical Statistics},
 number = {1},
 pages = {105--110},
 publisher = {Institute of Mathematical Statistics},
 title = {Conditional Expectation and Unbiased Sequential Estimation},
 urldate = {2025-10-22},
 volume = {18},
 year = {1947}
}

@Inbook{Rao1992,
author="Rao, C. Radhakrishna",
editor="Kotz, Samuel
and Johnson, Norman L.",
title="Information and the Accuracy Attainable in the Estimation of Statistical Parameters",
bookTitle="Breakthroughs in Statistics: Foundations and Basic Theory",
year="1992",
publisher="Springer New York",
address="New York, NY",
pages="235--247",
isbn="978-1-4612-0919-5",
doi="10.1007/978-1-4612-0919-5_16",
url="https://doi.org/10.1007/978-1-4612-0919-5_16"
}

@article{8a19a051-2be1-32f8-8b83-075ff1f85473,
 ISSN = {00063444, 14643510},
 URL = {http://www.jstor.org/stable/2337434},
 author = {George Casella and Christian P. Robert},
 journal = {Biometrika},
 number = {1},
 pages = {81--94},
 publisher = {[Oxford University Press, Biometrika Trust]},
 title = {Rao-Blackwellisation of Sampling Schemes},
 urldate = {2025-10-22},
 volume = {83},
 year = {1996}
}

@book{LehmannCasella1998,
  author    = {Lehmann, Erich L. and Casella, George},
  title     = {Theory of Point Estimation},
  edition   = {2nd},
  year      = {1998},
  publisher = {Springer},
  address   = {New York}
}

@article{SissonFan2007,
  title={A distance-based diagnostic for trans-dimensional Markov chains},
  author={S. A. Sisson and Y. Fan},
  journal={Statistics and Computin},
  year={2007},
  volume={17},
  pages={357-367},
  url={https://doi.org/10.1007/s11222-007-9025-z}
}

@ARTICLE{1992StaSc...7..457G,
       author = {{Gelman}, Andrew and {Rubin}, Donald B.},
        title = "{Inference from Iterative Simulation Using Multiple Sequences}",
      journal = {Statistical Science},
         year = 1992,
        month = jan,
       volume = {7},
        pages = {457-472},
          doi = {10.1214/ss/1177011136},
       adsurl = {https://ui.adsabs.harvard.edu/abs/1992StaSc...7..457G}
}

@book{gardiner2004handbook,
  address = {Berlin},
  author = {Gardiner, C. W.},
  edition = {Third},
  isbn = {3-540-20882-8},
  mrclass = {00A69 (60-01 60Hxx 60Jxx 82C31)},
  mrnumber = {2053476 (2004m:00008)},
  pages = {xviii+415},
  publisher = {Springer-Verlag},
  series = {Springer Series in Synergetics},
  title = {Handbook of stochastic methods for physics, chemistry and the
              natural sciences},
  volume = 13,
  year = 2004
}

@misc{colpi2024lisadefinitionstudyreport,
      title={LISA Definition Study Report}, 
      author={Monica Colpi and others},
      year={2024},
      eprint={2402.07571},
      archivePrefix={arXiv},
      primaryClass={astro-ph.CO},
      url={https://arxiv.org/abs/2402.07571}, 
}

@article{Green1995RJMCMC,
 ISSN = {00063444, 14643510},
 URL = {http://www.jstor.org/stable/2337340},
 author = {Peter J. Green},
 journal = {Biometrika},
 number = {4},
 pages = {711--732},
 publisher = {[Oxford University Press, Biometrika Trust]},
 title = {Reversible Jump Markov Chain Monte Carlo Computation and Bayesian Model Determination},
 urldate = {2025-10-30},
 volume = {82},
 year = {1995}
}

@article{gupta2001history,
author = {Gupta, Rameshwar D. and Richards, Donald St. P.},
title = {The History of the Dirichlet and Liouville Distributions},
journal = {International Statistical Review},
volume = {69},
number = {3},
pages = {433-446},
doi = {https://doi.org/10.1111/j.1751-5823.2001.tb00468.x},
url = {https://onlinelibrary.wiley.com/doi/abs/10.1111/j.1751-5823.2001.tb00468.x},
year = {2001}
}

@inproceedings{rasmussen2000infinite,
 author = {Rasmussen, Carl},
 booktitle = {Advances in Neural Information Processing Systems},
 editor    = {Solla, Sara A. and Leen, Todd K. and M{\"u}ller, Klaus-Robert},
 pages     = {554--560},
 publisher = {MIT Press},
 title = {The Infinite Gaussian Mixture Model},
 url = {https://proceedings.neurips.cc/paper_files/paper/1999/file/97d98119037c5b8a9663cb21fb8ebf47-Paper.pdf},
 volume = {12},
 year = {1999}
}

@article{metropolis1953equation,
  author    = {Metropolis, Nicholas and Rosenbluth, Arianna W. and Rosenbluth, Marshall N. and Teller, Augusta H. and Teller, Edward},
  title     = {Equation of state calculations by fast computing machines},
  journal   = {Journal of Chemical Physics},
  volume    = {21},
  number    = {6},
  pages     = {1087--1092},
  year      = {1953},
  doi       = {10.1063/1.1699114}
}

@article{hastings1970monte,
  author    = {Hastings, W. Keith},
  title     = {Monte Carlo sampling methods using Markov chains and their applications},
  journal   = {Biometrika},
  volume    = {57},
  number    = {1},
  pages     = {97--109},
  year      = {1970},
  doi       = {10.1093/biomet/57.1.97}
}

@inbook{neal2011mcmc,
  author    = {Neal, Radford M.},
  title     = {MCMC using Hamiltonian Dynamics},
  booktitle = {Handbook of Markov Chain Monte Carlo},
  editor    = {Brooks, Stephen and Gelman, Andrew and Jones, Galin and Meng, Xiao-Lin},
  pages     = {113--162},
  publisher = {Chapman \& Hall/CRC},
  address   = {New York},
  year      = {2011},
  doi       = {10.1201/b10905}
}

@ARTICLE{karamanis2021zeus,
       author = {{Karamanis}, Minas and {Beutler}, Florian and {Peacock}, John A.},
        title = "{zeus: a PYTHON implementation of ensemble slice sampling for efficient Bayesian parameter inference}",
      journal = {\mnras},
         year = 2021,
        month = dec,
       volume = {508},
       number = {3},
        pages = {3589-3603},
          doi = {10.1093/mnras/stab2867},
archivePrefix = {arXiv},
       eprint = {2105.03468},
 primaryClass = {astro-ph.IM},
       adsurl = {https://ui.adsabs.harvard.edu/abs/2021MNRAS.508.3589K}
}

@ARTICLE{karamanis2020ensemble,
       author = {{Karamanis}, Minas and {Beutler}, Florian},
        title = "{Ensemble Slice Sampling: Parallel, black-box and gradient-free inference for correlated \& multimodal distributions}",
      journal = {arXiv e-prints},
         year = 2020,
        month = feb,
          eid = {arXiv:2002.06212},
        pages = {arXiv:2002.06212},
          doi = {10.48550/arXiv.2002.06212},
archivePrefix = {arXiv},
       eprint = {2002.06212},
 primaryClass = {stat.ML},
       adsurl = {https://ui.adsabs.harvard.edu/abs/2020arXiv200206212K}
}

@article{geman1984stochastic,
  author    = {Geman, Stuart and Geman, Donald},
  title     = {Stochastic relaxation, Gibbs distributions, and the Bayesian restoration of images},
  journal   = {IEEE Transactions on Pattern Analysis and Machine Intelligence},
  volume    = {PAMI-6},
  number    = {6},
  pages     = {721--741},
  year      = {1984},
  doi       = {10.1109/TPAMI.1984.4767596}
}

@article{goodman2010ensemble,
  author    = {Goodman, Jonathan and Weare, Jonathan},
  title     = {Ensemble samplers with affine invariance},
  journal   = {Communications in Applied Mathematics and Computational Science},
  volume    = {5},
  number    = {1},
  pages     = {65--80},
  year      = {2010},
  doi       = {10.2140/camcos.2010.5.65}
}

@TechReport{repec:crs:wpaper:2001-29,
type={Working Papers},
institution={Center for Research in Economics and Statistics},
author={Olivier Cappe and Christian P, Robert and Tobias Ryden},
title={Reversible Jump MCMC Converging to Birth-and-Death MCMC and More General Continuous Time Samplers},
year={2001},
number={2001-29},
doi={None},
url={https://ideas.repec.org/p/crs/wpaper/2001-29.html},
}

@software{raynest,
       author = {{Del Pozzo}, W. and {Veitch}, J.},
        title = "{raynest: Parallel nested sampling based on ray}",
 howpublished = {Astrophysics Source Code Library, record ascl:2405.003},
         year = 2024,
        month = may,
          eid = {ascl:2405.003},
archivePrefix = {ascl},
       eprint = {2405.003},
       adsurl = {https://ui.adsabs.harvard.edu/abs/2024ascl.soft05003D}
}

@book{jaynes2003probability,
  author    = {Jaynes, E. T.},
  title     = {Probability Theory: The Logic of Science},
  publisher = {Cambridge University Press},
  address   = {Cambridge, UK},
  year      = {2003},
  editor    = {Bretthorst, G. Larry},
  isbn      = {9780521592710},
  doi       = {10.1017/CBO9780511790423}
}

@article{Cornish:2017vip,
    author = "Cornish, Neil and Robson, Travis",
    editor = "Giardini, Domencio and Jetzer, Philippe",
    title = "{Galactic binary science with the new LISA design}",
    eprint = "1703.09858",
    archivePrefix = "arXiv",
    primaryClass = "astro-ph.IM",
    doi = "10.1088/1742-6596/840/1/012024",
    journal = "J. Phys. Conf. Ser.",
    volume = "840",
    number = "1",
    pages = "012024",
    year = "2017"
}

@article{CarlinChib_product_space,
 ISSN = {00359246},
 URL = {http://www.jstor.org/stable/2346151},
 author = {Bradley P. Carlin and Siddhartha Chib},
 journal = {Journal of the Royal Statistical Society. Series B (Methodological)},
 number = {3},
 pages = {473--484},
 publisher = {[Royal Statistical Society, Oxford University Press]},
 title = {Bayesian Model Choice via Markov Chain Monte Carlo Methods},
 urldate = {2025-11-03},
 volume = {57},
 year = {1995}
}

@article{Cornish:2005qw,
    author = "Cornish, Neil J. and Crowder, Jeff",
    title = "{LISA data analysis using MCMC methods}",
    eprint = "gr-qc/0506059",
    archivePrefix = "arXiv",
    doi = "10.1103/PhysRevD.72.043005",
    journal = "Phys. Rev. D",
    volume = "72",
    pages = "043005",
    year = "2005"
}

@article{Vallisneri:2008ye,
    author = "Vallisneri, Michele",
    editor = "Lobo, Alberto and Sopuerta, Carlos F.",
    title = "{A LISA Data-Analysis Primer}",
    eprint = "0812.0751",
    archivePrefix = "arXiv",
    primaryClass = "gr-qc",
    doi = "10.1088/0264-9381/26/9/094024",
    journal = "Class. Quant. Grav.",
    volume = "26",
    pages = "094024",
    year = "2009"
}

@article{foremanmackey2013emcee,
  author    = {Foreman-Mackey, Daniel and Hogg, David W. and Lang, Dustin and Goodman, Jonathan},
  title     = {emcee: The {MCMC} Hammer},
  journal   = {Publications of the Astronomical Society of the Pacific},
  volume    = {125},
  number    = {925},
  pages     = {306--312},
  year      = {2013},
  doi       = {10.1086/670067},
  url       = {https://doi.org/10.1086/670067}
}

@article{maggiore2020einstein,
  author    = {Maggiore, Michele and Branchesi, Marica and Ghosh, Archisman and others},
  title     = {Science case for the Einstein Telescope},
  journal   = {Journal of Cosmology and Astroparticle Physics},
  year      = {2020},
  month     = {Mar},
  number    = {3},
  pages     = {050},
  doi       = {10.1088/1475-7516/2020/03/050},
  eprint    = {2003.01113},
  archivePrefix = {arXiv},
  primaryClass   = {astro-ph.IM}
}

@book{bookGelmanBayesianAnalysis3rd,
  address = {Boca Raton, Florida},
  author = {Gelman, Andrew and Carlin, John B. and Stern, Hal S. and Dunson, David B. and Vehtari, Akti and Rubin, Donald B.},
  edition = {Third},
  isbn = {9781439840955 1439840954},
  publisher = {CRC},
  refid = {966614951},
  series = {Chapman \& Hall/CRC Texts in Statistical Science Series},
  title = {Bayesian Data Analysis},
  url = {https://stat.columbia.edu/~gelman/book/},
  year = 2013
}

\end{document}